\def\prd{Phys. Rev. D}

\def\apj{Astrophys. J.}

\def\apjs{Astrophys. J.Suppl.}
\def\mnras{Mon. Not. R. Astr. Soc.}
\def\aap{Astr. Astrophys.}

\def \<{\langle}
\def \>{\rangle}

\newcommand{\ra}{\;\raise1.0pt\hbox{$'$}\hskip-6pt\partial\;}
\newcommand{\lo}{\;\overline{\raise1.0pt\hbox{$'$}\hskip-6pt\partial}\;}

\newcommand{\Abs}{\abstract}
\newcommand{\Ack}{\acknowledgments}
\newcommand{\mktt}{\maketitle}

\documentclass[a4paper,11pt]{article}
\pdfoutput=1

\usepackage{jcappub,graphicx,epsfig,natbib,color,times,bm,amsmath,multirow,hyperref}
\usepackage[ddmmyyyy,hhmmss]{datetime}

\begin{document}

\title{
General solutions of the leakage in integral transforms and applications to
the EB-leakage and detection of the cosmological gravitational wave background 
}

\author[a,b]{Hao Liu}\emailAdd{liuhao@nbi.dk}
\affiliation[a]{The Niels Bohr Institute \& Discovery Center, Blegdamsvej 17, DK-2100 Copenhagen, Denmark}
\affiliation[b]{Key Laboratory of Particle and Astrophysics, Institute of High Energy Physics, CAS, 19B YuQuan Road, Beijing, China}

\Abs{

For an orthogonal integral transform with complete dataset, any two components
are linearly independent; however, when some data points are missing, there is
going to be leakage from one component to another, which is referred to as the
``leakage in integral transforms'' in this work. A special case of this kind
of leakage is the EB-leakage in detection of the cosmological gravitational
wave background (CGWB). I first give the general solutions for all integral
transforms, prove that they are the best solutions, and then apply them to the
case of EB-leakage and detection of the CGWB. In the upcoming decade, most
likely, new cosmic microwave background (CMB) data are from ground/balloon
experiments, so they provide only partial sky coverage. Even in a fullsky
mission, due to the Galactic foreground, part of the sky is still unusable.
Within this context, the EB-leakage becomes inevitable. I show how to use the
general solutions to achieve the minimal error bars of the EB-leakage, and use
it to find out the maximum ability to detect the CGWB through CMB. The results
show that, when focusing on the tensor-to-scalar ratio $r$ (at a pivot scale
of 0.05 Mpc$^{-1}$), $1\%$ sky coverage ($f_{sky}=1\%$) is enough for a
$5\sigma$-detection of $r\ge 10^{-2}$, but is barely enough for $r=10^{-3}$.
If the target is to detect $r\sim 10^{-4}$ or $10^{-5}$, then $f_{sky}\ge
10\%$ is strongly recommended to enable a $5\sigma$-detection and to reserve
some room for other errors.

}


\mktt

\section{Introduction}\label{sec:intro}

The B-mode polarization of the CMB provides the most probable way of detecting
the CGWB, but the ability of detection is limited by the quality of foreground
removal, noise reduction, systematics control, delensing and EB-leakage
correction. In the coming decade, a practical constraint for detecting the
CGWB is that: Because all running experiments are ground based, for several
years there is going to be no new fullsky CMB data. Even when one tries to
combine the data of several ground experiments, there are still problems like
channel differences, systematics differences, observational time allocation,
etc. Therefore, in the near future, the detection of the CGWB through CMB has
to be done with a limited sky coverage.

When the sky is incomplete, there is going to be leakage from the much
stronger E-mode signal to the desired B-mode signal~\cite{2002PhRvD..65b3505L,
Bunn:2002df}, called the EB-leakage. This leakage will seriously contaminate
the primordial B-mode signal, so it must be corrected before detecting the
CGWB. It is reasonable to expect that, given an incomplete sky coverage, there
is going to be an unbeatable minimal error coming from the EB-leakage, which
will set an upper limit of the ability to detect the CGWB, even if all other
issues are perfectly solved. In~\cite{2018arXiv181104691L,
2019arXiv190400451L}, we presented the best blind estimate (BBE) of the
EB-leakage, but one problem still remains: when there is some reasonable prior
information, can we improve the estimation of the EB-leakage? If the answer is
yes, to what extent?

To answer this question, we should first give a very brief review of the
commonly used method of estimation in CMB science. Given a sky map
$\bm{P}=\{x_1,x_2,\cdots,x_n\}$ that can be described by a set of model
parameters $\bm{\Theta}=\{\theta_1,\theta_2,\cdots\,\theta_m\}$, a typical
parameter estimation problem is called a posterior estimation, which tries to
find $\bm{\Theta}$ that can maximize the conditional probability
$P(\bm{\Theta}|\bm{P})$ with given $\bm{P}$. It was clearly described
in~\cite{1997PhRvD..55.5895T}, given a sky map $\bm{P}$, how to design the
posterior best unbiased estimate (BUE) by using the Fisher information matrix.
Later in~\cite{2003ApJS..148..195V}, the estimation of $\bm{\Theta}$ was given
in a different way using the maximum likelihood approach. In
appendix~\ref{app:QML and WMAP MCL}, I provide a step-by-step proof that, for
a Gaussian isotropic CMB signal, the Fisher estimator and the maximum
likelihood estimator give identical results. Thus the problem of posterior
estimation in CMB is clear and sufficiently studied.

However, one thing has to be noticed: if some data points are missing, then
even a ``posterior estimation'' can not be 100\% posterior. There must be some
additional constraints/information about the missing data, otherwise the error
of estimates does not converge. For example, Gaussianity and isotropy of the
CMB signal are explicitly assumed in~\cite{1997PhRvD..55.5895T}. Below I
argue that, in addition to Gaussianity and isotropy, for the problem of
EB-leakage, the EE-spectrum can serve as perfect prior information:
\begin{enumerate}
\item \label{itm:prior 1} Currently in CMB experiments, the EE-spectrum is
known much better than the mysterious BB-spectrum. Especially, the Planck
mission already gave an excellent EE-spectrum by \emph{full sky}
surveys~\citep{2016A&A...594A...1P, 2016A&A...594A..11P}.
\item \label{itm:prior 2} Even if the EE-spectrum is slightly imperfect, there
is no problem to assume an ideal EE-spectrum to give an ideal lower limit of
the EB-leakage error bars, which is still very useful.
\item Practically, the estimation with prior information is not sensitive to
small variations of the EE-spectrum. Thus small imperfections are not really
important\footnote{According to eq.~(\ref{equ:final estimator}), the form of
the BUE is a matrix multiplication, and in the pixel domain, this effectively
means to estimate one pixel by the weighted sum of its neighboring pixels.
Therefore, to change the input EE-spectrum means to change the weighting
scheme. Because the two-point covariance of the CMB is linearly connected with
the CMB spectra (see e.g., Appendix A2 of ~\cite{2001PhRvD..64f3001T}), a
relatively small change of the input EE-spectrum means a relatively small
change of the weighting scheme. Furthermore, the change of the weighted sum is
normally much less than the change of the weighting scheme. Therefore, the BUE
result is not sensitive to small imperfection of the input EE-spectrum.
Especially, according to \citep{2016A&A...594A..11P}, the uncertainties of the
Planck EE power spectrum is already quite small. }.
\end{enumerate}

Therefore, with the best EE-spectrum given by Planck as the prior information,
it is expected that the BBE of the EB-leakage can be improved. In fact, the
solutions to this problem work not only for the EB-leakage, but also for all
kinds of integral transforms, thus this work is arranged as the following: I
start from the general case for all integral transforms, derive the general
solutions for both BBE and BUE, and give detailed proofs
(sections~\ref{sec:notations}--\ref{sec:masking leakage theorems}). Then the
solution is applied to the case of EB-leakage to give an estimation of the
maximum ability of detecting the CGWB through CMB (section~\ref{sec:tests}).
Finally, a brief discussion is given in section~\ref{sec:discuss}.

\section{Context and notations}\label{sec:notations}

To give a clear definition of ``leakage due to missing data in integral
transforms'', we should start from introducing the mathematical environment.

Let $\bm{\Omega}$ be an $n$-dimensional space, $\bm{p}=(x_1,x_2,\cdots,x_n)$
be one point in $\bm{\Omega}$, $\bm{\omega}$ be a subset of $\bm{\Omega}$
containing points $\{\bm{p}_1,\bm{p}_2,\cdots\}$, and $g=\{g_1(\bm{p}),
g_2(\bm{p}), \cdots\}_{\bm{p}\in\bm{\omega}}$ be a set of complete and
normalized basis functions defined on $\bm{\omega}$. Based on $g$, the
integral transform of a function $f(\bm{p})_{\bm{p}\in\bm{\omega}}$ (shortened
as $f(\bm{p})$ hereafter) is
\begin{eqnarray}\label{equ:ai}
a_i = \int_{\bm{\omega}} f(\bm{p})g_i^*(\bm{p})\,d\bm{p}.
\end{eqnarray}
Depending on the type of the integral transform, some conditions might be
required to ensure the convergence of the integral transform, e.g., the
Dirichlet conditions for Fourier transform. Normally, in the discrete case,
most such conditions can be ignored or at least weakened, which is very
convenient.

If the integral transforms are orthogonal, then we have
\begin{eqnarray}\label{equ:orth}
\int_{\bm{\omega}} g_i(\bm{p})g_j^*(\bm{p})\,d\bm{p} & = & 
\left\{
\begin{array}{lr} 
1 & (i=j) \\ 
0 & (i\ne j)
\end{array}
\right. .
\end{eqnarray}
This is also convenient, because it means $f(\bm{p})$ can be uniquely
decomposed into the summation of several components:
\begin{eqnarray}\label{equ:decomp}
f(\bm{p}) = \sum_i a_i g_i(\bm{p}) = \sum_i G_i(\bm{p}),
\end{eqnarray}
where $G_i(\bm{p})=a_i g_i(\bm{p})$ is the $i$-th component of $f(\bm{p})$.
Note that here the decomposition is written in a discrete form, whereas for
the continuous form, the summation becomes integration.

For convenience, the operation of extracting $G_i(\bm{p})$ from a function
$f(\bm{p})$ is shortened as
\begin{eqnarray}\label{equ:short gg}
\Psi_i(\bm{f})\Rightarrow G_i(\bm{p}).
\end{eqnarray}

When the data on some points are missing, the rest of the available points
forms a subset of $\bm{\omega}$ called $\bm{\omega}_1$. For convenience, this
is represented by a mask $\bm{M}(\bm{p})$ defined as
\begin{eqnarray}\label{equ:mask}
\bm{M}(\bm{p}) & = & 
\left\{
\begin{array}{lr} 
1 & (\bm{p}\in\bm{\omega}_1) \\ 
0 & (\bm{p}\notin\bm{\omega}_1)
\end{array}
\right. .
\end{eqnarray}
When the mask $\bm{M}(\bm{p})$ is present, the $i$-th component and the $j$-th
component of the integral transform are normally non-orthogonal. Thus the
$j$-th component derived with a mask will also receive a contribution from the
$i$-th component. This is called the $i$-to-$j$ leakage $\bm{L}_{ji}
(\bm{p}_1, \bm{p}_2, \cdots, \bm{q}_1,\bm{q}_2, \cdots)$, where
$\bm{p}_i\in\bm{\omega}_1$ and $\bm{q}_i\notin\bm{\omega}_1$ are the available
and unavailable points, respectively. For example, when $i$ represents the CMB
E-mode and $j$ represents the CMB B-mode, $\bm{L}_{ji}$ is called the
EB-leakage.

A precise estimation of $\bm{L}_{ji}$ cannot be achieved as it requires both
$\bm{p}_i$ and $\bm{q}_i$, however, if $\bm{L}_{ji}$ can be mathematically
decomposed into
\begin{equation}\label{equ:exp}
\bm{L}_{ji}(\bm{p}_1,\bm{p}_2,\cdots,\bm{q}_1,\bm{q}_2,\cdots) =
\bm{\mathcal{L}}_{ji}(\bm{p}_1,\bm{p}_2,\cdots) + 
\bm{\Delta}_{ji}(\bm{q}_1,\bm{q}_2,\cdots) +
\mathrm{const},
\end{equation}
where $\bm{\mathcal{L}}_{ji}$ and $\bm{\Delta}_{ji}$ depend only on the
available and unavailable points, respectively, then $\bm{\mathcal{L}}_{ji}$
is the best blind estimate, because any improvement of $\bm{\mathcal{L}}_{ji}$
requires additional information of the unavailable points, which is impossible
in the blind case. With the Taylor series expansion of $\bm{L}_{ji}$, one can
easily prove that, if eq.~(\ref{equ:exp}) exists, then $\bm{\mathcal{L}}_{ji}$
is unique. Therefore, if the BBE exists, then it is also unique (allowing a
trivial constant offset).

In a real experiment, it is possible that, although some points are missing,
one still knows some prior information about them, e.g., they are expected to
be Gaussian and isotropic. An estimation can be done either without or with
prior information $\bm{\mathcal{I}}$ of the missing points, called the blind
and non-blind estimations, respectively. For the non-blind estimation, let the
set of all $f(\bm{p})$ that satisfy the prior information $\bm{\mathcal{I}}$
be $\{f_i(\bm{p})\}_{\bm{\mathcal{I}}}$, then for convenience, I define the
following covariance matrices:
\begin{eqnarray}
\bm{C}_0(\bm{p},\bm{p}') &=& \< f(\bm{p}) 
	f(\bm{p}') \>_{\bm{\mathcal{I}}} \\ \nonumber
\bm{C}_1(\bm{p},\bm{p}') &=& \< \bm{L}_{ji}(\bm{p}) 
	\bm{\mathcal{L}}_{ji}(\bm{p}') \>_{\bm{\mathcal{I}}} \\ \nonumber
\bm{C}_2(\bm{p},\bm{p}') &=& \< \bm{\mathcal{L}}_{ji}(\bm{p}) 
	\bm{\mathcal{L}}_{ji}(\bm{p}') \>_{\bm{\mathcal{I}}}.
\end{eqnarray}

In the blind case, an important fact is that all the above covariance matrices
do not converge. However, with the prior information $\bm{\mathcal{I}}$, it is
possible for $\bm{C}_0(\bm{p},\bm{p}')$ to converge. In this work,
$\bm{\mathcal{I}}$ is called \textbf{regular} if $\bm{C}_0(\bm{p},\bm{p}')$
does converge and carry all information of $\bm{\mathcal{I}}$.

\section{The main results and proofs}\label{sec:masking leakage theorems}

With the above context and notation, the main results of this work are
summarized below, followed by detailed proofs:

\begin{enumerate}
\item \label{itm:best}In the blind case, the BBE of $\bm{L}_{ji}(\bm{p})$ is
unique, and is given by $\bm{\mathcal{L}}_{ji}(\bm{p}) =
\Psi_j(\bm{M}\bm{W}\Psi_i(\bm{M}\bm{f}))$, where $\bm{f}$ is the dataset,
$\bm{M}$ is the mask, and $\bm{W}$ is an optional window function.
\item \label{itm:causual}In the non-blind case, if the prior information
$\bm{\mathcal{I}}$ is regular, then the BUE of $\bm{L}_{ji}(\bm{p})$ is
$\bm{\mathcal{L}}^{\bm{\mathcal{I}}}_{ji}(\bm{p}) =
\bm{C}_1 \cdot\bm{C}_2^{-1}\cdot\bm{\mathcal{L}}_{ji}(\bm{p})$.
\end{enumerate}

\subsection{Proof of the blind case}\label{sub:proof blind}

First substitute eq.~(\ref{equ:ai}) into eq.~(\ref{equ:decomp}) to get the $i$-th
component of the integral transform:
\begin{eqnarray}\label{equ:get i}
G_i(\bm{p})&=& \int_{\bm{\omega}} f(\bm{p}')g_i^*(\bm{p}')g_i(\bm{p})\,d\bm{p}' \\ \nonumber
&=& \int_{\bm{\omega}} \mathcal{G}_i(\bm{p},\bm{p}') f(\bm{p}')\,d\bm{p}',
\end{eqnarray}
where $\mathcal{G}_i(\bm{p},\bm{p}') = g_i(\bm{p})g_i^*(\bm{p}')$ is the
real-space convolution kernel for getting the $i$-th component of the integral
transform. Use eq.~(\ref{equ:get i}) again to get the $j$-th component from
$G_i(\bm{p})$, which is the $i$-to-$j$ leakage $\bm{L}^0_{ji}(\bm{p})$ while
no data is missing:
\begin{eqnarray}
\bm{L}^0_{ji}(\bm{p}) &=& \int_{\bm{\omega}}
\mathcal{G}_j(\bm{p},\bm{p}') G_i(\bm{p}')\,d\bm{p}' \\ \nonumber 
&=& \int_{\bm{\omega}} \mathcal{G}_j(\bm{p},\bm{p}')
\left[\int_{\bm{\omega}} \mathcal{G}_i(\bm{p}',\bm{p}'')
f(\bm{p}'')\,d\bm{p}''\right]\,d\bm{p}' \\ \nonumber 
&=& \int_{\bm{\omega}}f(\bm{p}'')\mathcal{G}_{ji}(\bm{p},\bm{p}'')\,d\bm{p}'',
\end{eqnarray}
where $\mathcal{G}_{ji}(\bm{p},\bm{p}'') = \int_{\bm{\omega}}
\mathcal{G}_j(\bm{p},\bm{p}') \mathcal{G}_i(\bm{p}',\bm{p}'')\,d\bm{p}' $ is
the convolution kernel of the $i$-to-$j$ leakage. If the integral transform is
orthogonal, and the integral is done for the entire $\bm{\omega}$ (no data is
missing) and $j\ne i$, then $\mathcal{G}_{ji} (\bm{p},\bm{p}'') = 0$ and
$\bm{L}^0_{ji}(\bm{p}) = 0$, i.e., there is no leakage.

When the data on some points are missing as described by eq.~(\ref{equ:mask}),
the $i$-to-$j$ leakage $\bm{L}_{ji}(\bm{p})$ is given by getting the $j$-th
component from the real $G_i(\bm{p}')$ with a mask. Using the notation in
eq.~(\ref{equ:short gg}), this is written as
\begin{eqnarray}
\bm{L}_{ji}(\bm{p})=\Psi_j(\bm{M}\Psi_i(\bm{f})),
\end{eqnarray}
whose integral form is
\begin{eqnarray}\label{equ:leakage full}
\bm{L}_{ji}(\bm{p})&=& \int_{\bm{\omega}} \mathcal{G}_j(\bm{p},\bm{p'}) 
	\left[\bm{M}(\bm{p}')G_i(\bm{p}')\right] \,d\bm{p}' \\ \nonumber
&=& \int_{\bm{\omega}}  \mathcal{G}_j(\bm{p},\bm{p'})\bm{M}(\bm{p}')
	\left[\int_{\bm{\omega}} \mathcal{G}_i(\bm{p}',\bm{p}'')
f(\bm{p}'')\,d\bm{p}''\right]\,d\bm{p}' \\ \nonumber
&=& \int_{\bm{\omega}}  f(\bm{p}'') 
	\mathcal{G}_{ji}(\bm{p},\bm{p}'')_{\bm{\omega_1}}\,d\bm{p}'',
\end{eqnarray}
where
\begin{eqnarray}\label{equ:kernel with mask}
\mathcal{G}_{ji}(\bm{p},\bm{p}'')_{\bm{\omega_1}} &=& \int_{\bm{\omega}}
\mathcal{G}_j(\bm{p},\bm{p'}) \mathcal{G}_i(\bm{p}',\bm{p}'') \bm{M}(\bm{p}')\,d\bm{p}'
\end{eqnarray}
is a \emph{fixed} convolution kernel, which is normally non-zero, so the
leakage $\bm{L}_{ji}(\bm{p})$ is normally non-zero.

Eq.~(\ref{equ:leakage full}) shows that the true leakage term is the convolution
of the kernel $\mathcal{G}_{ji}(\bm{p},\bm{p}'')_{\bm{\omega_1}}$ with
$f(\bm{p}'')$ on the entire $\bm{\omega}$, which cannot be done for the
missing part of the data. In reality, this integral can only be done for
$\bm{\omega}_1$, which gives
\begin{eqnarray}\label{equ:max leakage}
\bm{\mathcal{L}}_{ji}(\bm{p})= \int_{\bm{\omega}}  f(\bm{p}'')\bm{M}(\bm{p}'')
\mathcal{G}_{ji}(\bm{p},\bm{p}'')_{\bm{\omega}_1}\,d\bm{p}'',
\end{eqnarray}
and the error $\bm{\Delta}_{ji}(\bm{p})$ is
\begin{eqnarray}\label{equ:error leakage}
\bm{\Delta}_{ji}(\bm{p}) = \bm{L}_{ji}(\bm{p})-\bm{\mathcal{L}}_{ji}(\bm{p}) =
\int_{\bm{\omega}}  f(\bm{p}'')[1-\bm{M}(\bm{p}'')]
\mathcal{G}_{ji}(\bm{p},\bm{p}'')_{\bm{\omega}_1}\,d\bm{p}''.
\end{eqnarray}
$\bm{\mathcal{L}}_{ji}(\bm{p})$ and $\bm{\Delta}_{ji}(\bm{p})$ fully satisfy
eq.~(\ref{equ:exp}), thus $\bm{\mathcal{L}}_{ji}(\bm{p})$ is the unique BBE.

Calculation of eq.~(\ref{equ:max leakage}) is difficult, because it is normally
computationally intensive to obtain $\mathcal{G}_{ji}(\bm{p},
\bm{p}'')_{\bm{\omega}_1}$. A much faster way of doing the same thing is to
calculate $\bm{\mathcal{L}}_{ji}^{fast}(\bm{p})$ instead, which requires only
to change the order of integration as follows:
\begin{eqnarray}\label{equ:fast}
\bm{\mathcal{L}}_{ji}(\bm{p})
&=& \int_{\bm{\omega}} f(\bm{p}'')\bm{M(\bm{p}'')}  
	\mathcal{G}_{ji}(\bm{p},\bm{p}'')_{\bm{\omega}_1}\,d\bm{p}''  \\ \nonumber 
&=& \int_{\bm{\omega}} f(\bm{p}'')\bm{M(\bm{p}'')} \left[\int_{\bm{\omega}}
	\mathcal{G}_j(\bm{p},\bm{p'}) \mathcal{G}_i(\bm{p}',\bm{p}'') \bm{M}(\bm{p}')
	\,d\bm{p}'\right]\,d\bm{p}''   \\ \nonumber 
&=& \int_{\bm{\omega}} \mathcal{G}_j(\bm{p},\bm{p}')\bm{M(\bm{p}')} \left[
	\int_{\bm{\omega}} \mathcal{G}_i(\bm{p}',\bm{p}'') f(\bm{p}'') \bm{M(\bm{p}'')} 
	\,d\bm{p}'' \right] \,d\bm{p}'    \\ \nonumber 
&=& \bm{\mathcal{L}}_{ji}^{fast}(\bm{p}) .
\end{eqnarray}
Based on the above equation and using the notation in eq.~(\ref{equ:short gg}),
the analytic form of the unique BBE that is easy to calculate is:
\begin{eqnarray}\label{equ:short}
\bm{\mathcal{L}}_{ji}(\bm{p})\equiv \bm{\mathcal{L}}_{ji}^{fast}(\bm{p}) = 
	\Psi_j(\bm{M}\Psi_i(\bm{M}\bm{f})).
\end{eqnarray}
Eq.~(\ref{equ:short}) is easy to calculate because it can be done by standard
forward-backward integral transforms. Therefore, the BBE can be easily
obtained as long as a fast algorithm of the integral transform is available,
e.g., the fast Fourier transforms (FFT). This is extremely convenient, and is
exactly the idea of the recycling method used in~\cite{2018arXiv181104691L} to
correct the EB-leakage for detecting the CGWB.

It is also possible to use an extra window function $\bm{W}(\bm{p})$ that
smooths the edge of the region, which is the conventional way of reducing the
leakage. It is important to note that the window function is defined only in
the available region, and it can never change the unavailable region to
available. Therefore, using the window function is equivalent to replacing
$\bm{M}(\bm{p}')$ by $\bm{M}(\bm{p}')\bm{W}(\bm{p}')$ in eq.~(\ref{equ:leakage
full}) and eq.~(\ref{equ:fast}) (without touching $\bm{M}(\bm{p}'')$).
Consequently, eq.~(\ref{equ:short}) becomes
\begin{eqnarray}\label{equ:short with window}
\bm{\mathcal{L}}_{ji}(\bm{p})\equiv \bm{ \mathcal{L} }_{ji}^{fast}(\bm{p}) = 
\Psi_j(\bm{MW}\Psi_i(\bm{M}\bm{f})).
\end{eqnarray}

\subsection{Proof of the non-blind case}\label{sub:proof prior}

Now we can start to prove the conclusion for the non-blind case. For
convenience, the true leakage $\bm{L}_{ji}(\bm{p})$ is shortened as a column
vector $\bm{L}$. Similarly, $\bm{\mathcal{L}}_{ji}(\bm{p})$ is shortened as
column vector $\bm{\mathcal{L}}$ of the same size. Hence the general form of
$\bm{L}$ as a function of $\bm{\mathcal{L}}$ is
\begin{eqnarray}
\bm{L} = \widetilde{\bm{L}}(\bm{\mathcal{L}}) + 
	\bm{\Lambda}(\bm{\mathcal{L}})+\bm{\Delta},
\end{eqnarray}
where $\widetilde{\bm{L}}(\bm{\mathcal{L}})$ is the linear part whose second
and higher derivatives are zero, and $\bm{\Lambda}(\bm{\mathcal{L}})$ is the
non-linear part whose first derivative is zero. $\bm{\Delta}$ is the error,
which is not a function of $\bm{\mathcal{L}}$ and is statistically
uncorrelated with $\widetilde{\bm{L}}(\bm{\mathcal{L}})$ and
$\bm{\Lambda}(\bm{\mathcal{L}})$. The matrix form of the above equation is
\begin{eqnarray}\label{equ:assumed form}
\bm{L} = \bm{m}\cdot\bm{\mathcal{L}} +
\bm{\Lambda}(\bm{\mathcal{L}}) + \bm{\Delta},
\end{eqnarray}
where $\bm{m}$ is a constant coupling matrix. 

If the prior information $\bm{\mathcal{I}}$ is regular, then $\bm{C}_0$ is
converged and constant. Using eq.~(\ref{equ:leakage full}--\ref{equ:max
leakage}), we get
\begin{eqnarray}\label{equ:cov ext}
\bm{C}_1
&=& \left\<\int_{\bm{\omega}}  f(\bm{p}') \mathcal{G}_{ji}(\bm{p},\bm{p}')_{\bm{\omega_1}} 
	f(\bm{p}''') \mathcal{G}_{ji}(\bm{p}'',\bm{p}''')_{\bm{\omega_1}} 
	\bm{M}(\bm{p}''')\,d\bm{p}'d\bm{p}'''\right\>_{\bm{\mathcal{I}}} \\ \nonumber
&=& \int_{\bm{\omega}}  \mathcal{G}_{ji}(\bm{p},\bm{p}')_{\bm{\omega_1}} 
	 \mathcal{G}_{ji}(\bm{p}'',\bm{p}''')_{\bm{\omega_1}} \bm{M}(\bm{p}''')
	 \left\<f(\bm{p}')f(\bm{p}''')\right\>_{\bm{\mathcal{I}}}
	 \,d\bm{p}'d\bm{p}''' \\ \nonumber 
&=& \int_{\bm{\omega}}  \mathcal{G}_{ji}(\bm{p},\bm{p}')_{\bm{\omega_1}} 
	 \mathcal{G}_{ji}(\bm{p}'',\bm{p}''')_{\bm{\omega_1}} \bm{M}(\bm{p}''')
	 \bm{C}_0(\bm{p}',\bm{p}''')\,d\bm{p}'d\bm{p}''' \\ \nonumber
&=& \mathrm{const}\,.
\end{eqnarray}
Similarly, $\bm{C}_2$ is also constant. Using eq.~(\ref{equ:assumed form}) to
recalculate the covariance matrix gives
\begin{eqnarray}\label{equ:re-cov}
\bm{C}_1 &=&\<\bm{L}\bm{\mathcal{L}^T}\>_{\bm{\mathcal{I}}} \\ \nonumber
&=& \bm{m}\cdot\<\bm{\mathcal{L}}\bm{\mathcal{L}}^T\>_{\bm{\mathcal{I}}} +
\<\bm{\Lambda}(\bm{\mathcal{L}})\bm{\mathcal{L}}^T\>_{\bm{\mathcal{I}}} + 
\<\bm{\Delta}\bm{\mathcal{L}}^T\>_{\bm{\mathcal{I}}} \\ \nonumber
&=& \bm{m}\cdot\bm{C}_2 ,
\end{eqnarray}
thus $\bm{m}=\bm{C}_1\cdot\bm{C}_2^{-1}$ is a constant matrix, and
\begin{eqnarray}\label{equ:for taylor}
\widetilde{\bm{L}} = \bm{m} \cdot \bm{\mathcal{L}} = 
\bm{C}_1\cdot\bm{C}_2^{-1}\cdot\bm{\mathcal{L}}.
\end{eqnarray}
With the well known theory of least-square fitting, one can easily prove that
eq.~(\ref{equ:for taylor}) is exactly the unbiased estimate with minimal
variance (see Appendix~\ref{app:least square} for more details). Therefore,
the BUE with regular prior information $\bm{\mathcal{I}}$ is
\begin{eqnarray}\label{equ:final estimator}
\bm{\mathcal{L}}^{\bm{\mathcal{I}}}_{ji}(\bm{p})=\bm{C}_1 \cdot
\bm{C}_2^{-1}\cdot\bm{\mathcal{L}}_{ji}(\bm{p}).
\end{eqnarray}

\subsection{Property of the error}\label{sub:error property}
In this section, the properties of the error of estimation are discussed,
which gives important information about when we should apply the leakage
estimation.

Starting from eq.~(\ref{equ:get i}), with mask $\bm{M}(\bm{p}')$, we get the
corrupted $i$-th component $\widetilde{G}_i(\bm{p}) $ as
\begin{eqnarray}\label{equ:error}
\widetilde{G}_i(\bm{p}) &=& \int_{\bm{\omega}} \mathcal{G}_i(\bm{p},\bm{p}') 
	f(\bm{p}')\bm{M}(\bm{p}')\,d\bm{p}' \\ \nonumber
&=& \int_{\bm{\omega}} \mathcal{G}_i(\bm{p},\bm{p}') \bm{M}(\bm{p}')
	\left[ \sum_j{G_j(\bm{p}')} \right] \,d\bm{p}' \\ \nonumber
&=& \int_{\bm{\omega}} \mathcal{G}_i(\bm{p},\bm{p}') \bm{M}(\bm{p}')
	\left[ G_i(\bm{p}') + \sum_{j\ne i}{G_j(\bm{p}')} \right] \,d\bm{p}' \\ \nonumber
&=& \int_{\bm{\omega}} \mathcal{G}_{ii}(\bm{p},\bm{p}')_{\bm{\omega_1}}f(\bm{p}')\,d\bm{p}' + 
	\int_{\bm{\omega}} \mathcal{G}_{i\bar{i}}(\bm{p},\bm{p}')_{\bm{\omega_1}}
	f(\bm{p}')\,d\bm{p}' \\ \nonumber
&=& \bm{L}_{ii}(\bm{p}) + \bm{L}_{i\bar{i}}(\bm{p}),
\end{eqnarray}
where $\bar{i}$ means the combination of all the $j\ne i$ components, so
$\bm{L}_{ii}(\bm{p})$ is the $i$-to-$i$ leakage, and
$\bm{L}_{i\bar{i}}(\bm{p})$ is the combined cross leakage from all non-$i$
components. In analogy to a square matrix, they are similar to the diagonal
and off-diagonal terms.

For a set of given input data and given mask, $\widetilde{G}_i(\bm{p})$ is
known and fixed, thus the errors of estimating $\bm{L}_{ii}(\bm{p})$ and
$\bm{L}_{i\bar{i}}(\bm{p})$ satisfy
\begin{eqnarray}\label{equ:error1}
\Delta\bm{L}_{ii}(\bm{p}) + \Delta \bm{L}_{i\bar{i}}(\bm{p})=0,
\end{eqnarray}
which means one can only try to optimize one estimator, and then the error of
the other term is automatically fixed. In practice, if the prior information
$\mathcal{I}$ shows that the $i$-th component is very weak, then apparently
$\Delta \bm{L}_{i\bar{i}}(\bm{p})$ will be dominating. In this case, the
overall best estimation is given by eq.~(\ref{equ:final estimator}).

For example, a standard integral transform of the polarized signal on the
sphere includes two modes: the E-mode and the B-mode. Each of them has the
same number of members. If the polarized signal is the CMB, then the prior
expectation shows that all B-mode components are much weaker than the E-mode
components, thus eq.~(\ref{equ:final estimator}) is the best choice of
correcting the E-to-B leakage.

\section{Application: the maximum ability to detect the CGWB through the Cosmic
Microwave Background with incomplete sky coverage}\label{sec:tests}

As mentioned in the introduction, in the coming decade, all available CMB
experiments are ground-based, which can hardly provide full sky coverage. In
this case, detection of the primordial B-mode will contain an unbeatable
minimal error due to the EB-leakage, even if everything else is done
perfectly. This sets an absolute constraint on the ability to detect the CGWB
through CMB for each choice of the sky coverage. Apparently, to find this
constraint, we should use the BUE with prior information $\mathcal{I}$ in
eq.~(\ref{equ:final estimator}), with $\mathcal{I}$ being the best available
EE-spectrum.

\subsection{Comparison of the BBE and the BUE}\label{sub:compare bue bbe}

\begin{figure}
  \centering
  \includegraphics[width=0.48\textwidth]{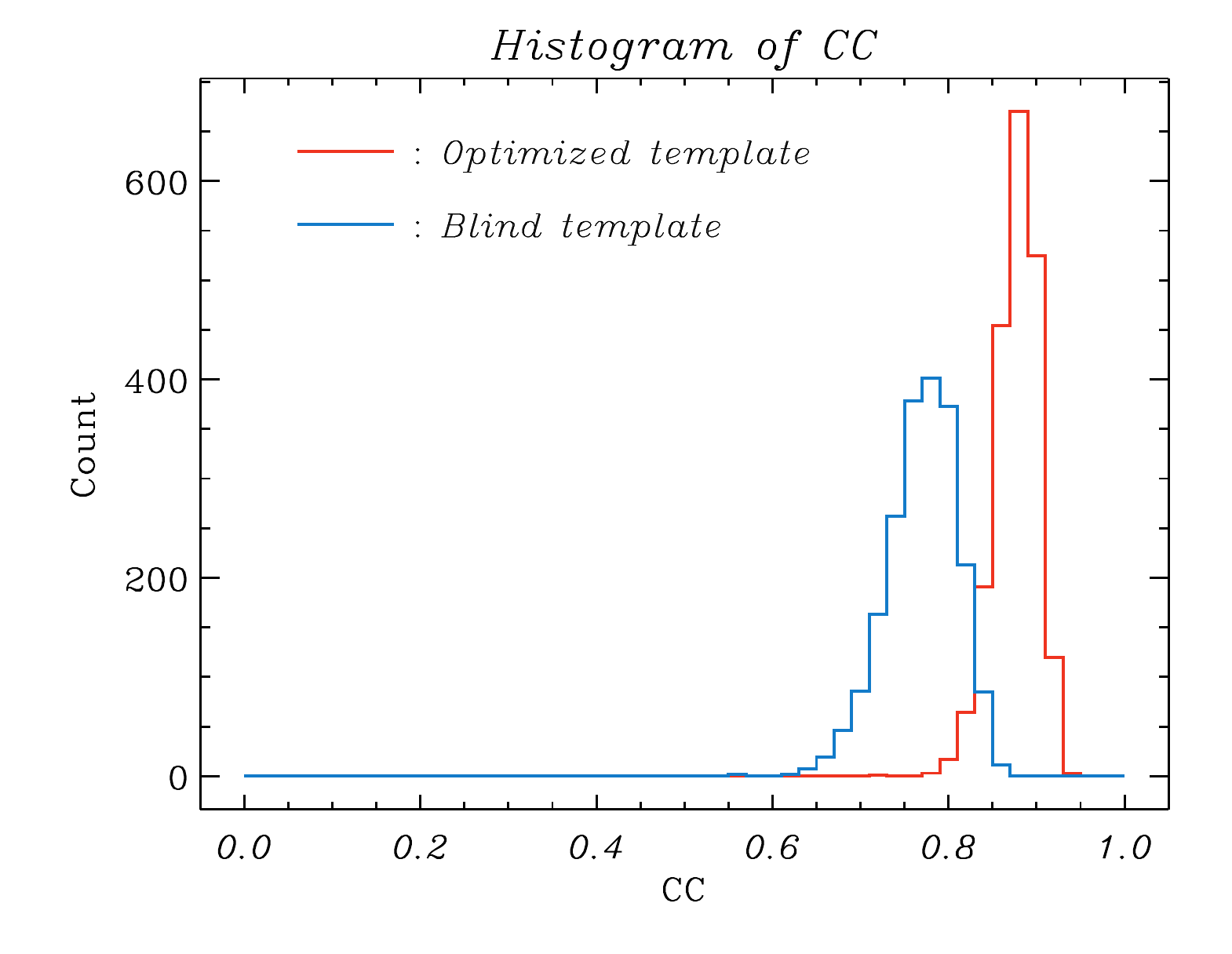}
  \includegraphics[width=0.48\textwidth]{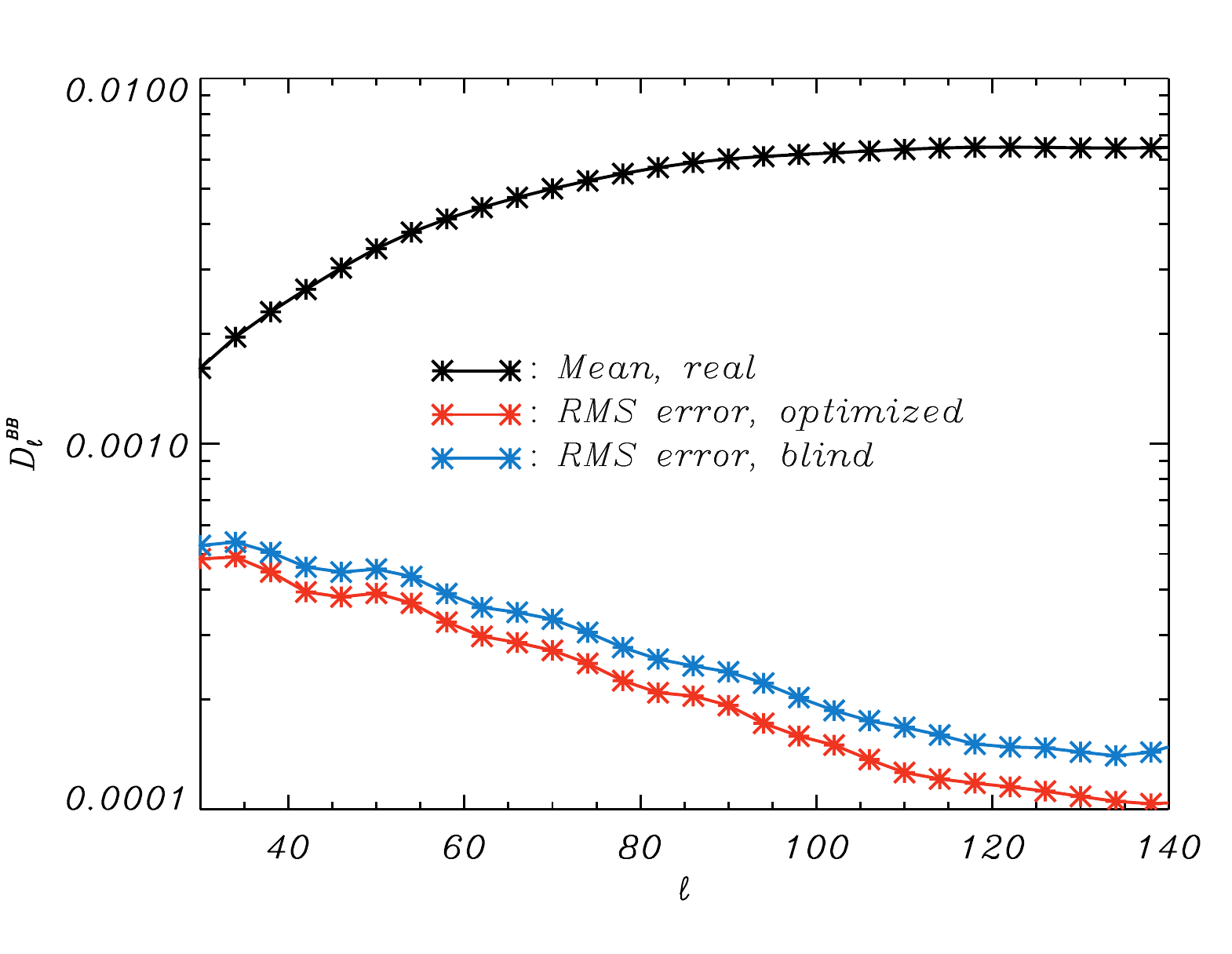}
  \caption{\emph{Left}: Comparison of the similarity between the real and
  fixed B-maps using the Pearson Cross Correlation coefficients (CC). Red for
  the BUE, and blue for the BBE, represented by the recycling method
  in~\citep{2018arXiv181104691L}. Apparently the BUE gives better similarity
  to the real B-map. \emph{Right}: Comparison of the BB-spectrum error of the
  BUE and the recycling method, where $D_{\ell}=\ell(\ell+1)C_{\ell}/(2\pi)$.
  The BUE gives roughly 30\% lower error bars around $\ell=100$. }
  \label{fig:cmp BUE}
\end{figure}

Now we proceed to testing and comparing the BBE and the BUE given in
section~\ref{sec:masking leakage theorems} with practical calculations. We
start from the best fit Planck 2018 CMB spectra to generate simulated CMB
maps. Because the BUE is extremely time consuming\footnote{The most time
consuming part of the BUE is to operate the two covariance matrices. The size
of the covariance matrices (number of rows) scales as $n\propto f_{sky}
N_{side}^2$, and the operation of matrix multiplication and inversion scales
as $n^3$, thus the time cost of the BUE scales as $f_{sky}^3 N_{side}^6$. },
the simulated CMB maps are generated only at $N_{side}=64$ with $r=0.05$, and
a disk mask of $f_{sky}=0.01$ is used in this test. For each simulated map,
the BBE and the BUE are used to estimate the resulting B-maps, respectively,
and compared the results with the real B-map (derived from the known fullsky
map) within the available region. We first compare their similarities by the
Pearson Cross Correlation (CC) coefficients, as shown in the left panel of
Figure~\ref{fig:cmp BUE}. Evidently, the result of the BUE (red line) has
higher CC coefficients (better similarity) with the real B-map than the BBE
does. In the right panel, I compare the BB-spectrum error of the BUE and the
BBE. Note that in the test here, the focus is on the error of the EB-leakage
estimation, thus the BB-spectra are calculated from the masked real B-map and
the BUE/BBE B-maps directly, without reconstruction of the fullsky spectra
(this will be considered later). The uncertainties of the BUE/BBE methods are
computed as the rms differences between the real B-map spectrum and the
BUE/BBE B-map spectra. The result shows that, when running at the same
resolution, the BUE helps to reduce the error around $\ell\approx100$ by
roughly 30\%.

Practically, as pointed out by~\cite{Bunn:2002df}, oversampling can help to
reduce the leakage due to pixelization, and a similar effect was seen even for
the temperature case, as shown in figures~4 and 6 of
~\cite{2014MNRAS.440..957M}. Because it is much easier to increase the
resolution for the BBE than for the BUE, practically, running the BBE at a
much higher resolution can effectively give a comparable result to running the
BUE at a relatively lower resolution. Thus running the BBE at increased
resolution is likely the most practical and promising way of correcting the
EB-leakage at a high resolution.

\subsection{Estimation of the maximum ability}

As discussed in section~\ref{sec:intro}, the maximum ability to detect the
CGWB through the Cosmic Microwave Background with incomplete sky coverage can
be found out by the BUE of the EB-leakage with proper prior information. The
details of the idea are as follows: when only part of the sky is available,
the minimal error of EB-leakage is given by the BUE in eq.~(\ref{equ:final
estimator}), and the minimal error of recovering the full sky BB-spectrum is
given by the Fisher estimator~\citep{1997PhRvD..55.5895T, 2001PhRvD..64f3001T,
2002ApJ...567....2H, 2009ApJ...697..258J, 2009MNRAS.400..463G,
2014MNRAS.440..957M}. Combining these two minimal errors gives the maximum
ability of detecting CGWB through CMB when all other things are assumed to be
perfect, except only the sky coverage is incomplete.

In practice, both the BUE in eq.~(\ref{equ:final estimator}) and the Fisher
estimator are extremely time consuming. For a real estimation of the
EB-leakage and the B-mode spectrum, one has to deal with the BUEs, but in this
section, the main interest is on the amplitude of the error, thus it is
possible to save a lot of time by simplifying the calculation, as to be
illustrated below.

\begin{figure}
  \centering
  \includegraphics[width=0.48\textwidth]{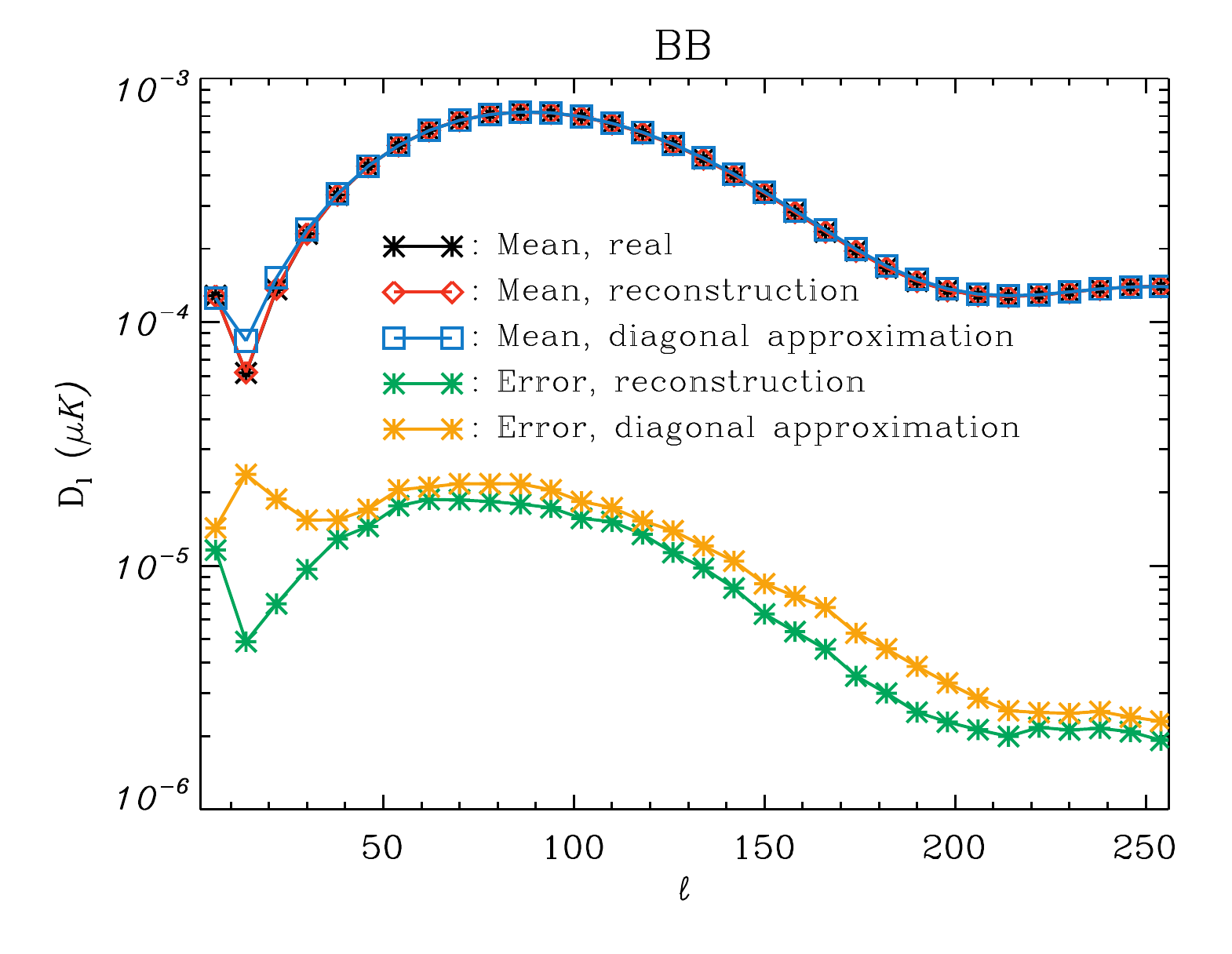}
  \caption{Comparison of the errors of the pseudo-$C_{\ell}$ estimator (green)
  and its diagonal approximation (yellow) when reconstructing the fullsky
  BB-spectrum (no lensing) from the real B-map with partial sky coverage.
  Generally speaking, at $\ell\approx100$, both of them are unbiased. The
  pseudo-$C_{\ell}$ estimator is evidently better, but the ability to reduce
  the error is only by 20--30\%. }
  \label{fig:pcl and diag}
\end{figure}

We know from Figure~\ref{fig:cmp BUE} that the BUE only gives 20--30\%
improvement compared to the recycling method used
in~\cite{2018arXiv181104691L, 2019arXiv190400451L}, and the latter is
sufficiently fast. Meanwhile, it was shown by Figure 6
of~\cite{2014MNRAS.440..957M} that for $\ell\approx100$, the error of the
fisher estimator is no less than 50\% of the pseudo-$C_{\ell}$ method. By
investigating the pseudo-$C_{\ell}$ method, it is further discovered in
Figure~\ref{fig:pcl and diag} that its error at $\ell\approx100$ is roughly
20\% lower than the one given by its diagonal approximation, which means to
divide the cut sky spectrum by a constant factor $k = \left[\sum_{\bm{n}}
W^2(\bm{n})\right]/N_{pix}$, where $W(\bm{n})$ is the mask/apodization
function. The diagonal approximation is acceptable for small scales like
$\ell>50$, but is bad at larger scales (lower $\ell$). Because the diagonal
approximation of the pseudo-$C_{\ell}$ method and the recycling method are
sufficiently fast, the minimal error $\Delta(\ell)_{\rm{min}}$ can be
efficiently calculated as\footnote{Again, it should be emphasized that these
simplified approaches work only for estimating the amplitude of error. For a
real reconstruction task, one has to go back to a full implementation of the
BUEs.}
\begin{eqnarray}
\Delta(\ell)_{\rm{min}} &\approx& \sqrt{\left(\frac{\Delta_1(\ell)}{1.3k}\right)^2 
+ \left(\frac{\Delta_2(\ell)}{2.5}\right)^2 + \Delta_c^2(\ell)},
\end{eqnarray}
where 100 simulations are run for each test in figure~\ref{fig:test
r5}--\ref{fig:test r2}. $\Delta_1(\ell)$ is the 1$\sigma$ error of the
EB-leakage correction using the recycling method, as shown by the blue line in
figure~\ref{fig:cmp BUE}. As one can see in figure~\ref{fig:cmp BUE}, a
dedicated correction using the BUE can reduce $\Delta_1(\ell)$ roughly by
factor 1.3. However, the difference between the BUE B-map and the real B-map
(the pixel domain error) is non-Gaussian and non-isotropic, thus
$\Delta_1(\ell)$ cannot benefit from the later fisher estimation of the
fullsky spectrum, and will only passively receive the impact of fisher
estimation. Because the leading order approximation of the fisher estimation
(for $\ell$ around 100) is to multiply with $1/k$, the contribution of the
residual EB-leakage error after applying the fisher estimator can be roughly
estimated as $\Delta_1(\ell)/(1.3k)$. $\Delta_2(\ell)$ is the 1$\sigma$ error
of recovering the full sky BB-spectrum using diagonal approximation on the
real B-map with partial sky coverage (e.g., the yellow line in
figure~\ref{fig:pcl and diag}). Similarly, by using the dedicated fisher
estimator, $\Delta_2(\ell)$ can be suppressed by factor 2.5\footnote{The exact
factor of improvement by using the fisher estimator will vary according to the
sky fraction and the multipole range, thus the factor 2.5 here is a rough
estimate. } at $\ell\approx100$. The relative strength of $\Delta_1(\ell)$ and
$\Delta_2(\ell)$ is determined by the level of $r$ and the sky fraction. When
only $r$ decreases, $\Delta_1(\ell)$ will not change, because it is dominated
by the EE-spectrum; however, $\Delta_2(\ell)$ will decrease, because for the
fisher estimator, the amplitude of error roughly follows the amplitude of the
true spectrum. When only $f_{sky}$ increases, both $\Delta_1(\ell)$ and
$\Delta_2(\ell)$ will decrease, because when more sky becomes available, all
errors will decrease\footnote{This is based on the assumption that the noise
level is a constant. For a real experiment, it is possible that, by decreasing
the sky fraction, more observations per pixel are obtained, and the error
could become lower. }. For example, at the recombination peak
($\ell\simeq100$), when $r=10^{-2}$ and $f_{sky}=10\%$, the term with
$\Delta_2(\ell)$ is dominating; when $r=10^{-3}$ and $f_{sky}=2\%$, the two
terms are comparable; but when $r=10^{-4}$ and $f_{sky}=2\%$, the term with
$\Delta_1(\ell)$ is dominating. Lastly, $\Delta_c(\ell)$ is the cosmic
variance, which is intrinsic and fixed. Therefore, $\Delta(\ell)_{\rm{min}}$
is a rough estimate of the minimal error compared to the theoretical
BB-spectrum, which represents the maximum ability to detect the CMB B-mode.

Amongst various shapes of the mask, the disc mask has best symmetry, which
helps to alleviate the EB-leakage. Thus disc masks with cosine apodizations
are used for tests. Let's consider the cases when the disc masks have
$f_{sky}=$ 1\%, 3\%, 5\%, 7\%, 10\% and 20\%; and the apodization parameters
are $a$ = 0.1, 0.3, $\cdots$, 0.9, where $a$ is the ratio between the
apodization width and the disc radius, so high $a$-value means more aggressive
apodization. The results are presented in figures~\ref{fig:test
r5}--\ref{fig:test r2} for $r=10^{-5}$, $10^{-4}$, $10^{-3}$ and ${10^{-2}}$,
respectively. Note that in order to focus on a $5\sigma$ detection, the errors
(colored lines) are amplified by factor 5, so by a simple comparison with the
theoretical BB-spectrum (black line), one can easily see whether or not a
$5\sigma$ detection is possible. Meanwhile, since an aggressive apodization
will significantly reduce the signal-to-noise ratio, a detection is promising
only when most of the colored lines are lower than the black line, so less
aggressive apodizations are allowed.

The main results of this section are given in figures~\ref{fig:test
r5}--\ref{fig:test r2}. Additionally, for convenience of reading, based on
these figures, brief comments are given for the detectability of $r$ in
Table~\ref{tab:det} for various values of $f_{sky}$, e.g., the BICEP2
observation region has $f_{sky}\approx 1.2\%$. The meaning of the words in
Table~\ref{tab:det} are listed below:
\begin{enumerate}
\item ``Impossible'': The specified $f_{sky}$ can hardly satisfy the
requirement of detection.
\item ``Barely'': The specified $f_{sky}$ can marginally satisfy the
requirement of detection, but there is little room for other errors.
\item ``Possible'': The specified $f_{sky}$ can satisfy the requirement of
detection, and there is also room for other errors.
\item ``Hopeful'': The specified $f_{sky}$ can satisfy the requirement of
detection, and there is considerable room for other errors.
\end{enumerate}

\begin{table}
 \centering
 \begin{tabular}{|l|l|l|l|l|} \hline
 			   & $r=10^{-5}$ & $r=10^{-4}$	& $r=10^{-3}$ & $r=10^{-2}$ \\ \hline
$f_{sky}=0.01$ & Impossible	 & Impossible	& Barely	  & Possible 	\\ \hline
$f_{sky}=0.03$ & Barely		 & Barely		& Possible	  & Hopeful		\\ \hline
$f_{sky}=0.05$ & Barely		 & Possible		& Hopeful	  & Hopeful		\\ \hline
$f_{sky}=0.07$ & Barely		 & Hopeful		& Hopeful	  & Hopeful		\\ \hline
$f_{sky}=0.10$ & Possible	 & Hopeful		& Hopeful	  & Hopeful		\\ \hline
$f_{sky}=0.20$ & Hopeful	 & Hopeful		& Hopeful	  & Hopeful		\\ \hline
 \end{tabular}
 \caption{ Rough comments on the possibility of detecting the CGWB at
 $5\sigma$ significance with various mask sizes. Based on
 figures~\ref{fig:test r5}--\ref{fig:test r2}. Note that perfect foreground
 removal, noise reduction, systematics control and delensing are assumed, thus
 this table shows the ultimate ability of detection.}
 \label{tab:det}
\end{table}

\begin{figure}
  \centering
  \includegraphics[width=0.32\textwidth]{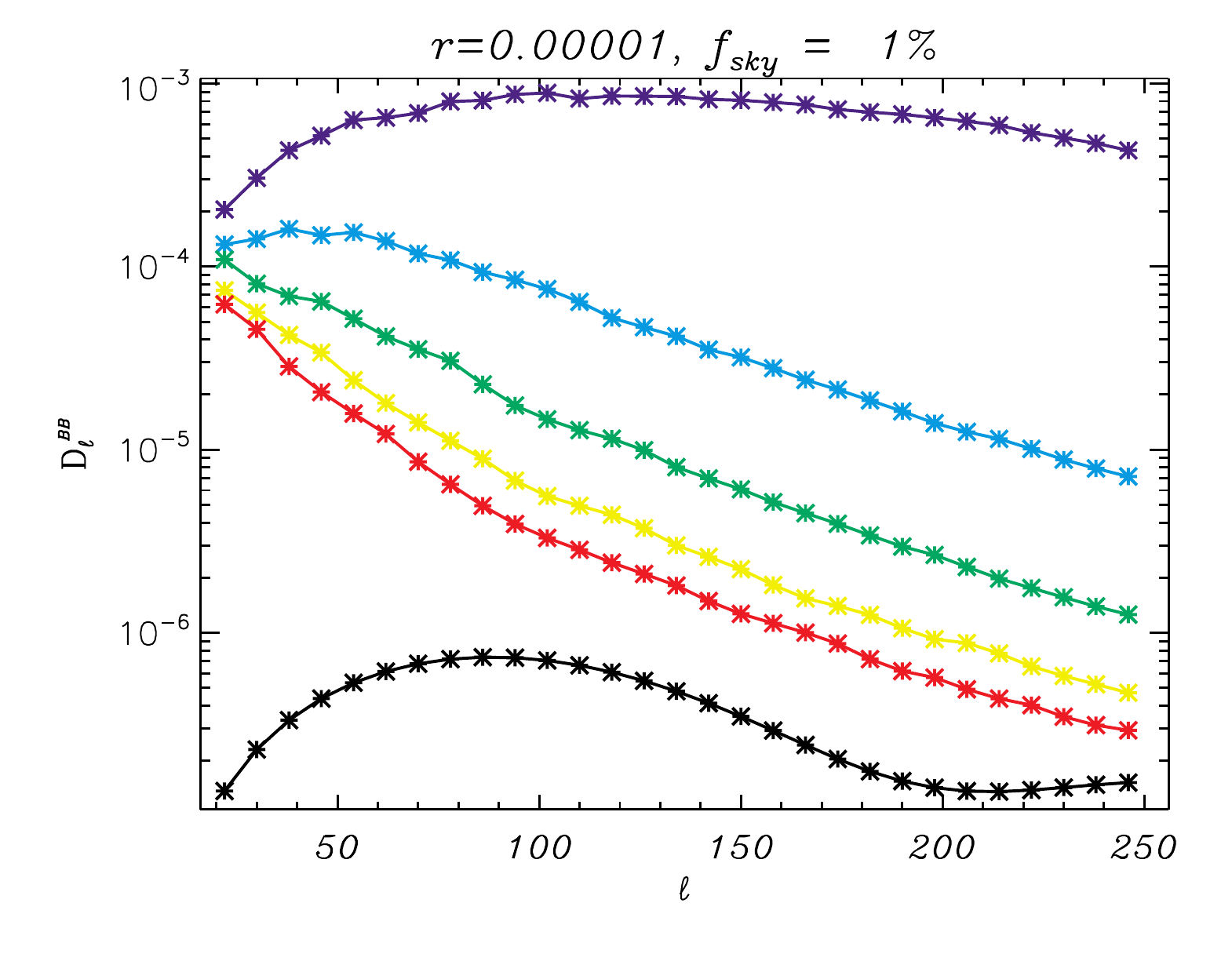}
  \includegraphics[width=0.32\textwidth]{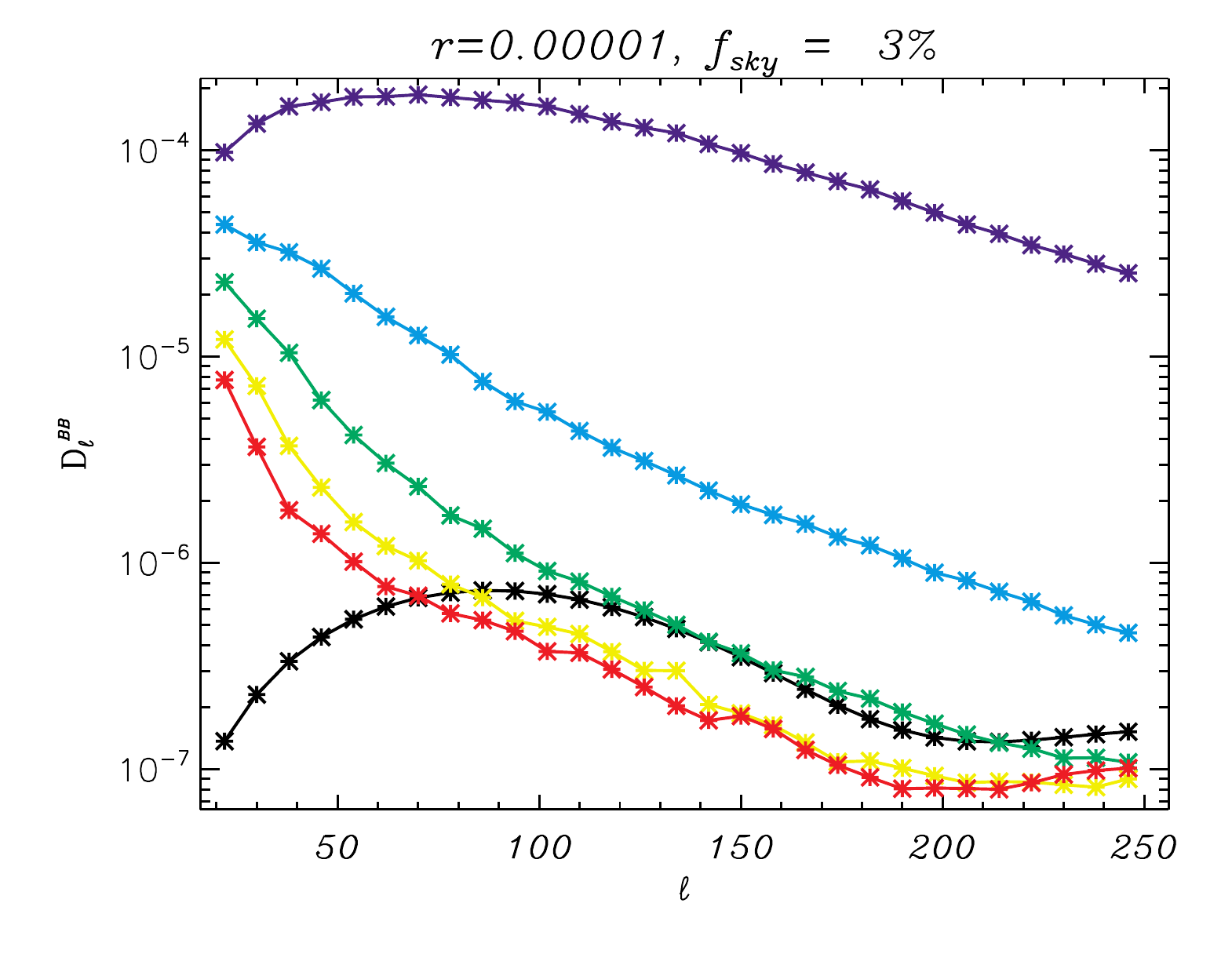}
  \includegraphics[width=0.32\textwidth]{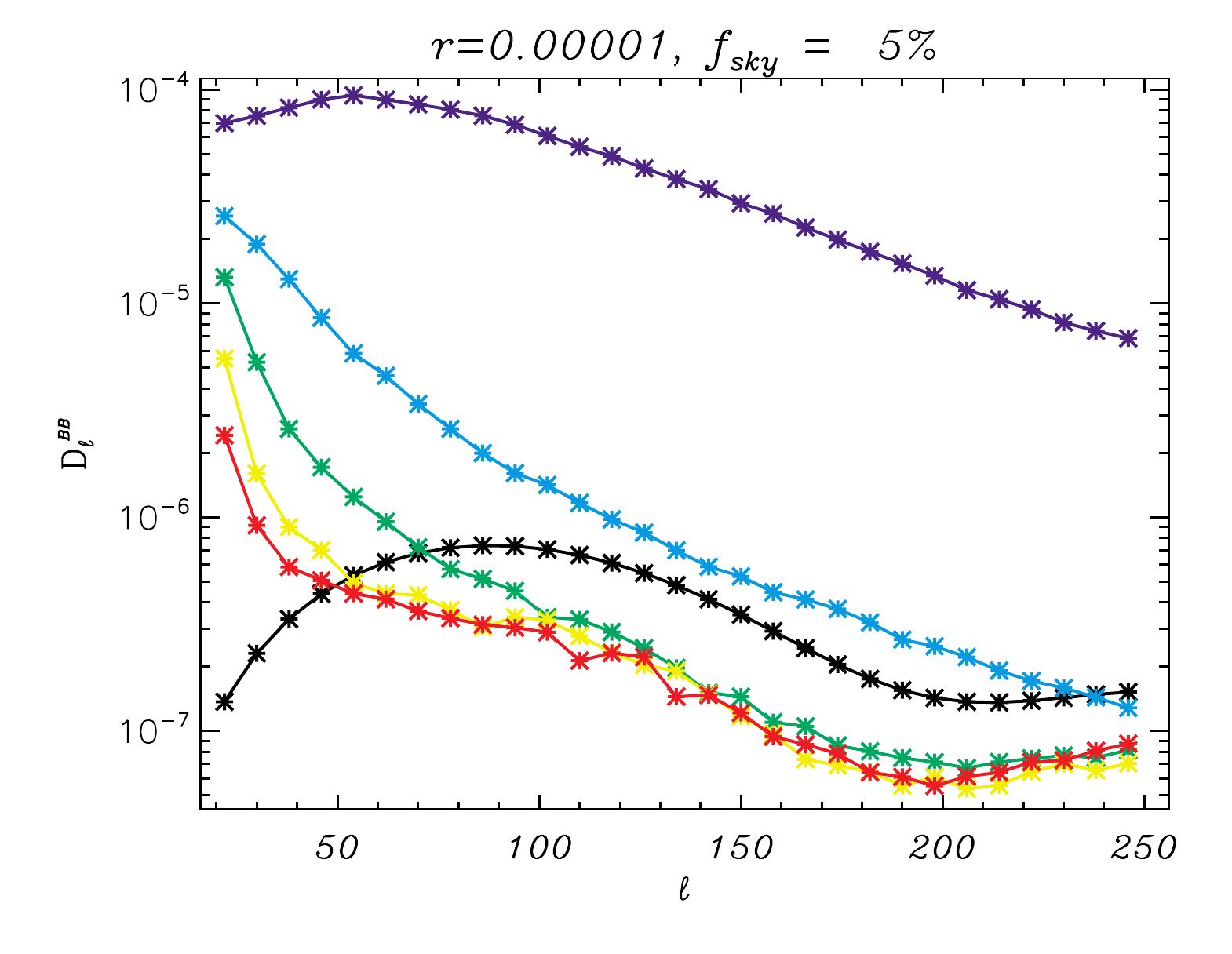}

  \includegraphics[width=0.32\textwidth]{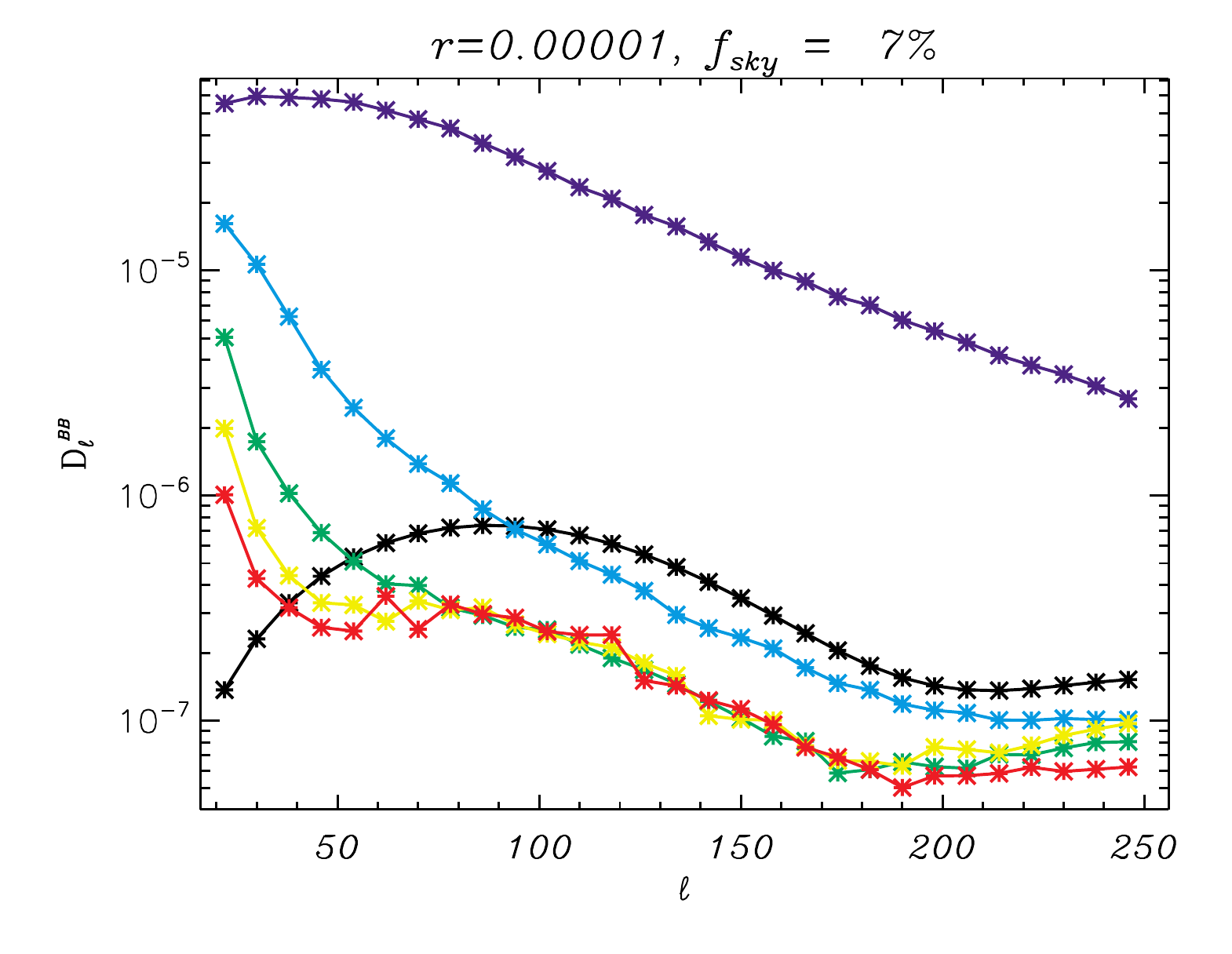}
  \includegraphics[width=0.32\textwidth]{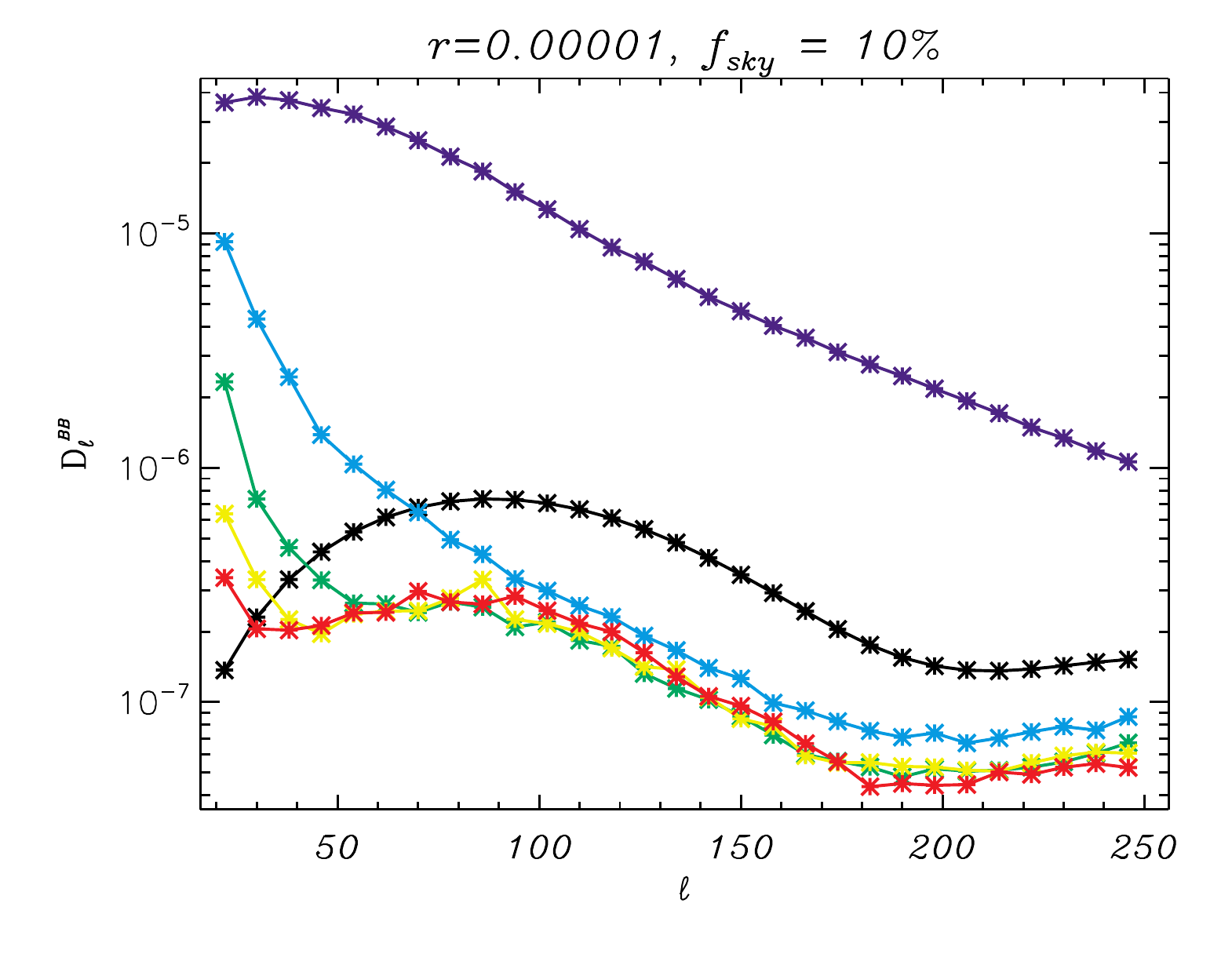}
  \includegraphics[width=0.32\textwidth]{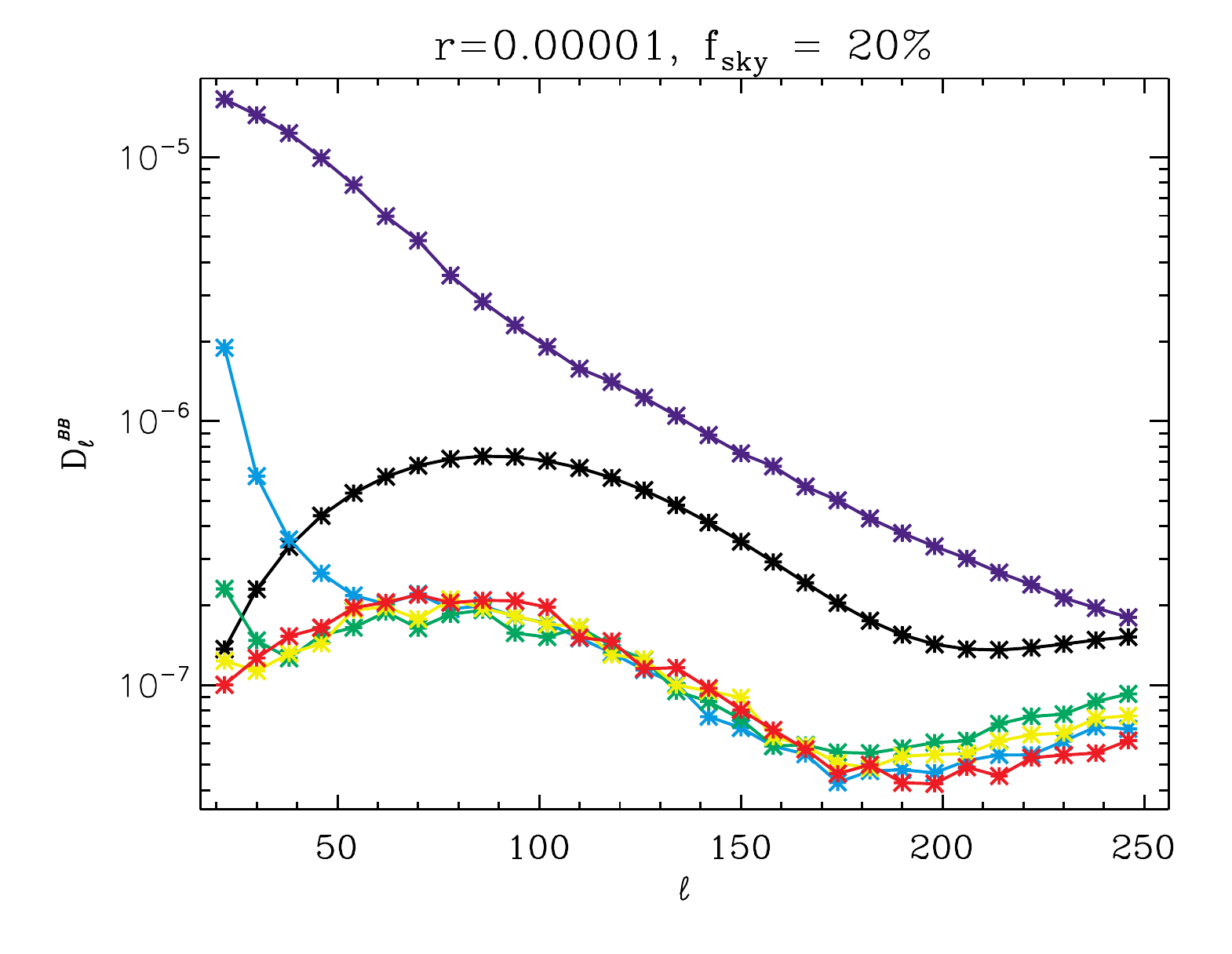}
  \caption{The lower limits of the error of detecting the CMB B-mode (colored
  lines). Note that they are amplified by factor 5 to show the threshold of
  $5\sigma$ detection. The colors corresponds to 10\% -- 90\% (deep purple to
  red) apodizations. The theoretical BB-spectrum of $r=10^{-5}$ without
  lensing is shown by the black line for comparison. From top-left to
  bottom-right: $f_{sky}=$ 0.01, 0.03, 0.05, 0.07, 0.10 and 0.20. Perfect
  foreground removal, noise reduction, systematics control and delensing are
  assumed, so this is the absolutely unavoidable error attached to each choice
  of mask size and apodization strength. }
  \label{fig:test r5}
\end{figure}

\begin{figure}
  \centering
  \includegraphics[width=0.32\textwidth]{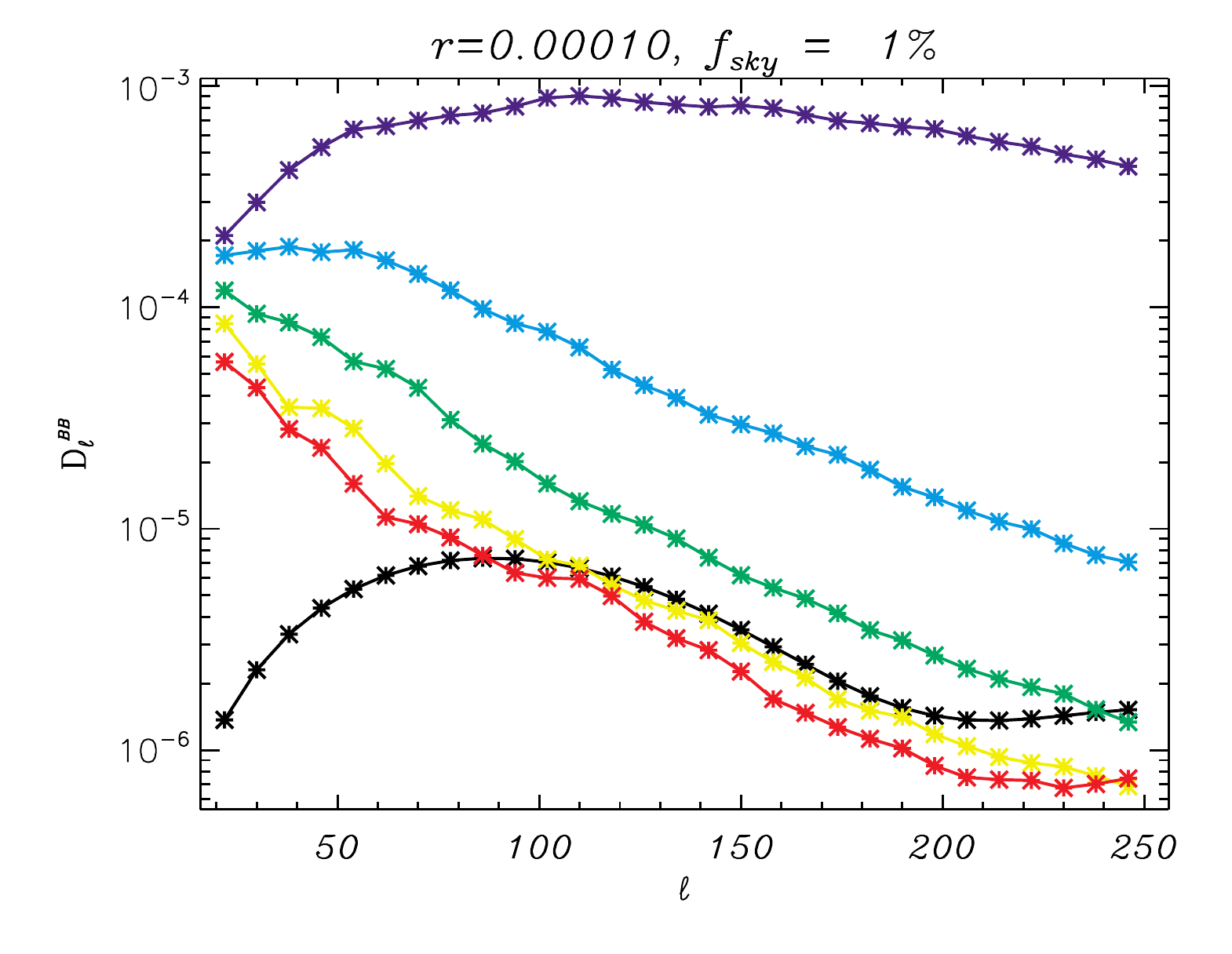}
  \includegraphics[width=0.32\textwidth]{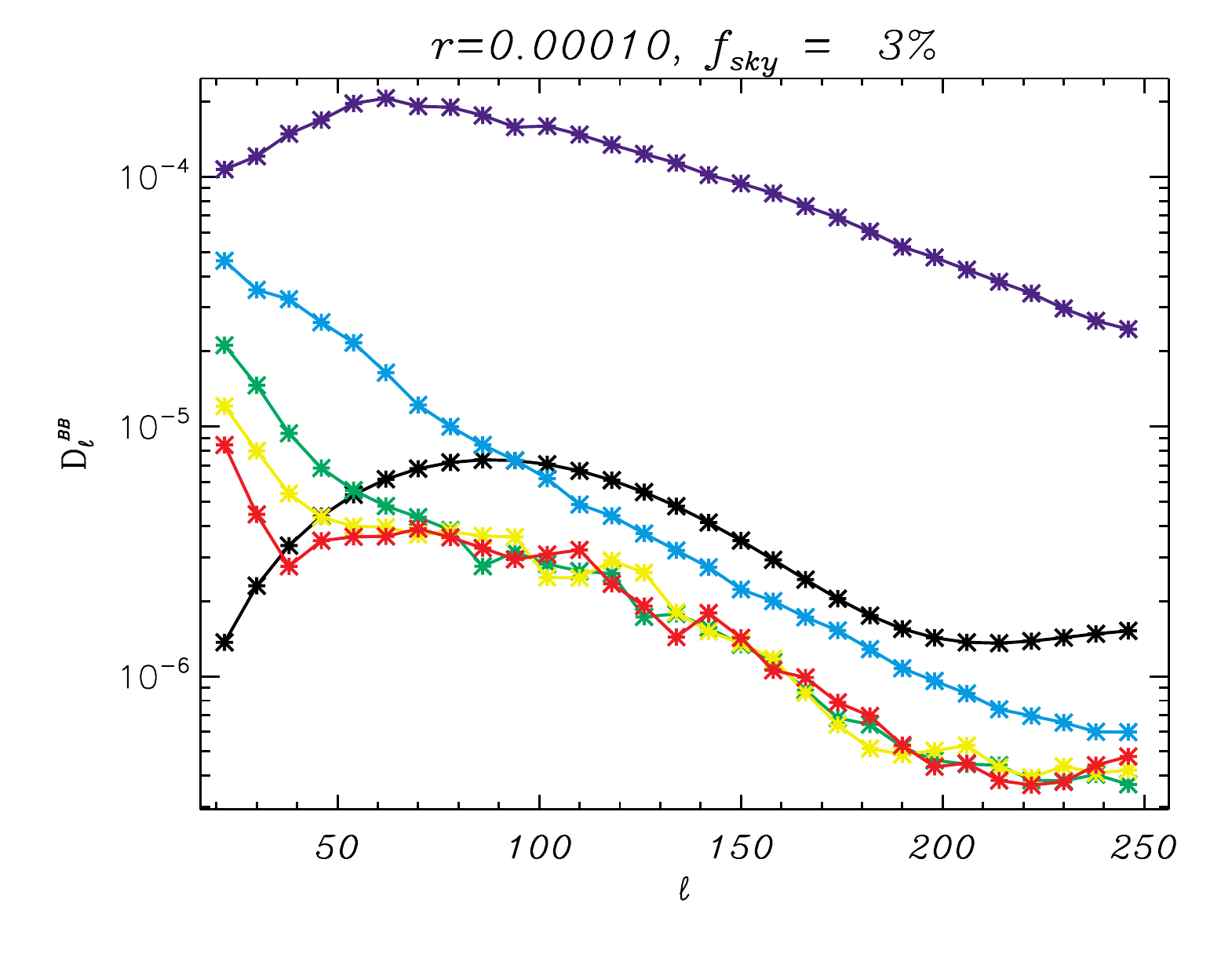}
  \includegraphics[width=0.32\textwidth]{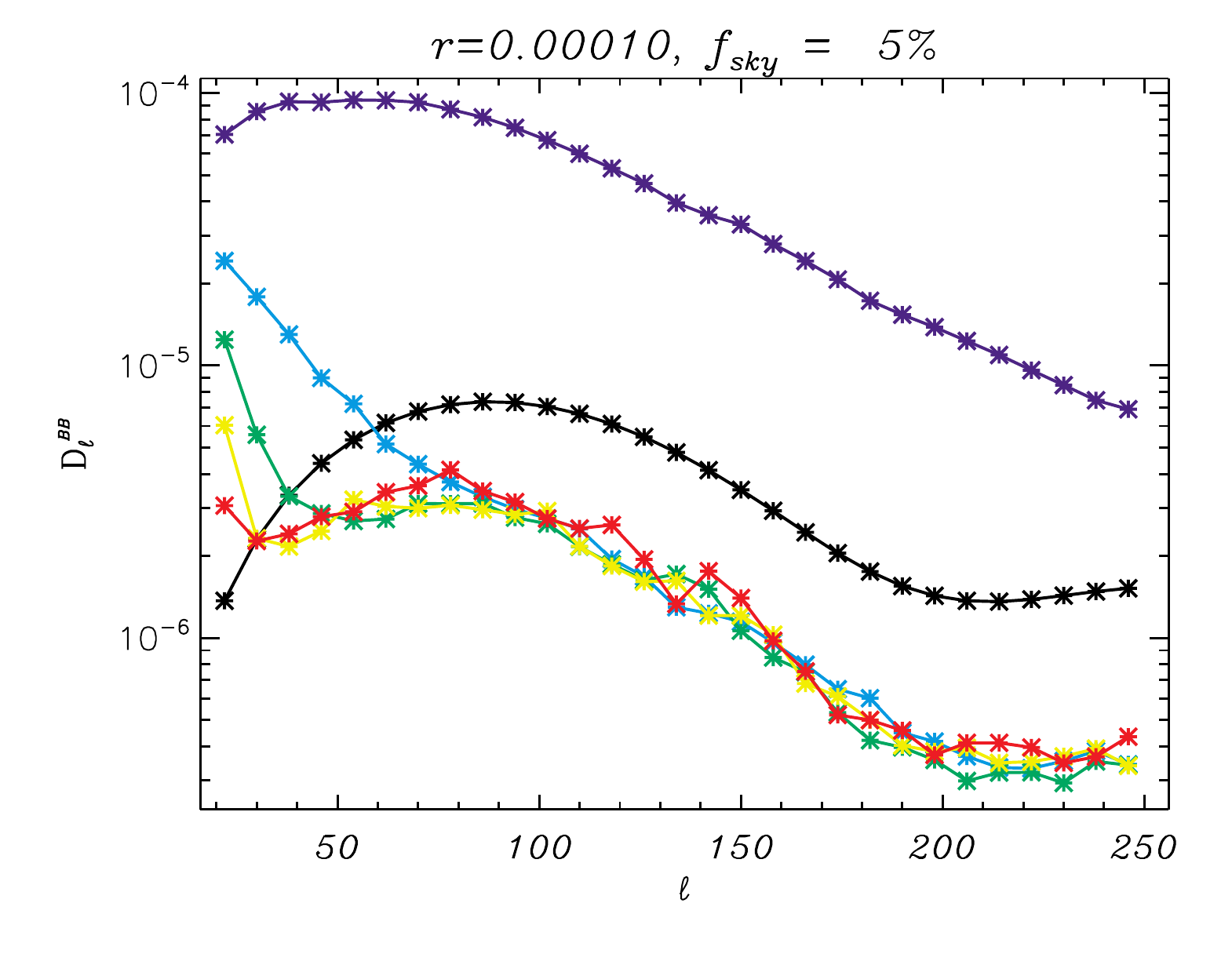}

  \includegraphics[width=0.32\textwidth]{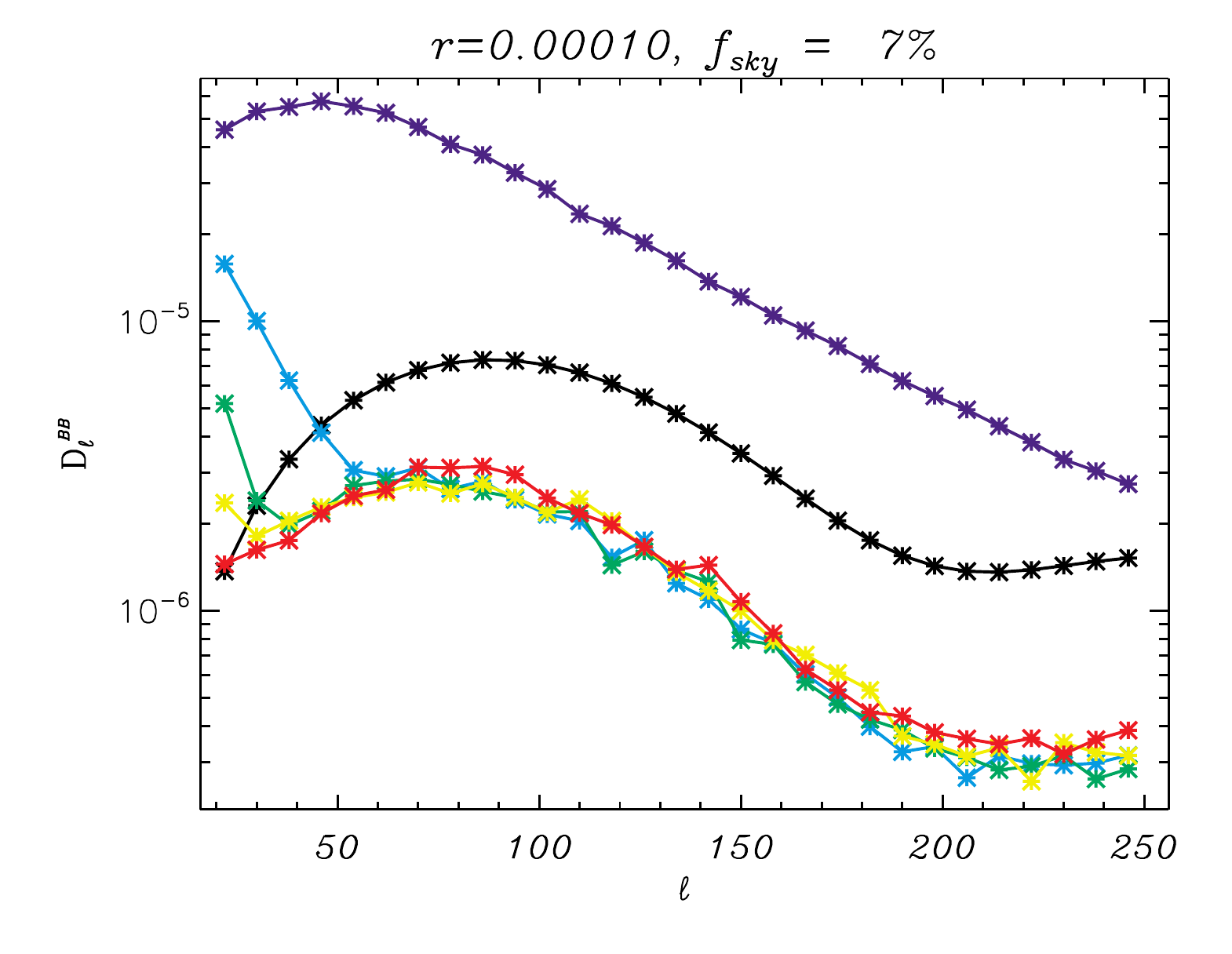}
  \includegraphics[width=0.32\textwidth]{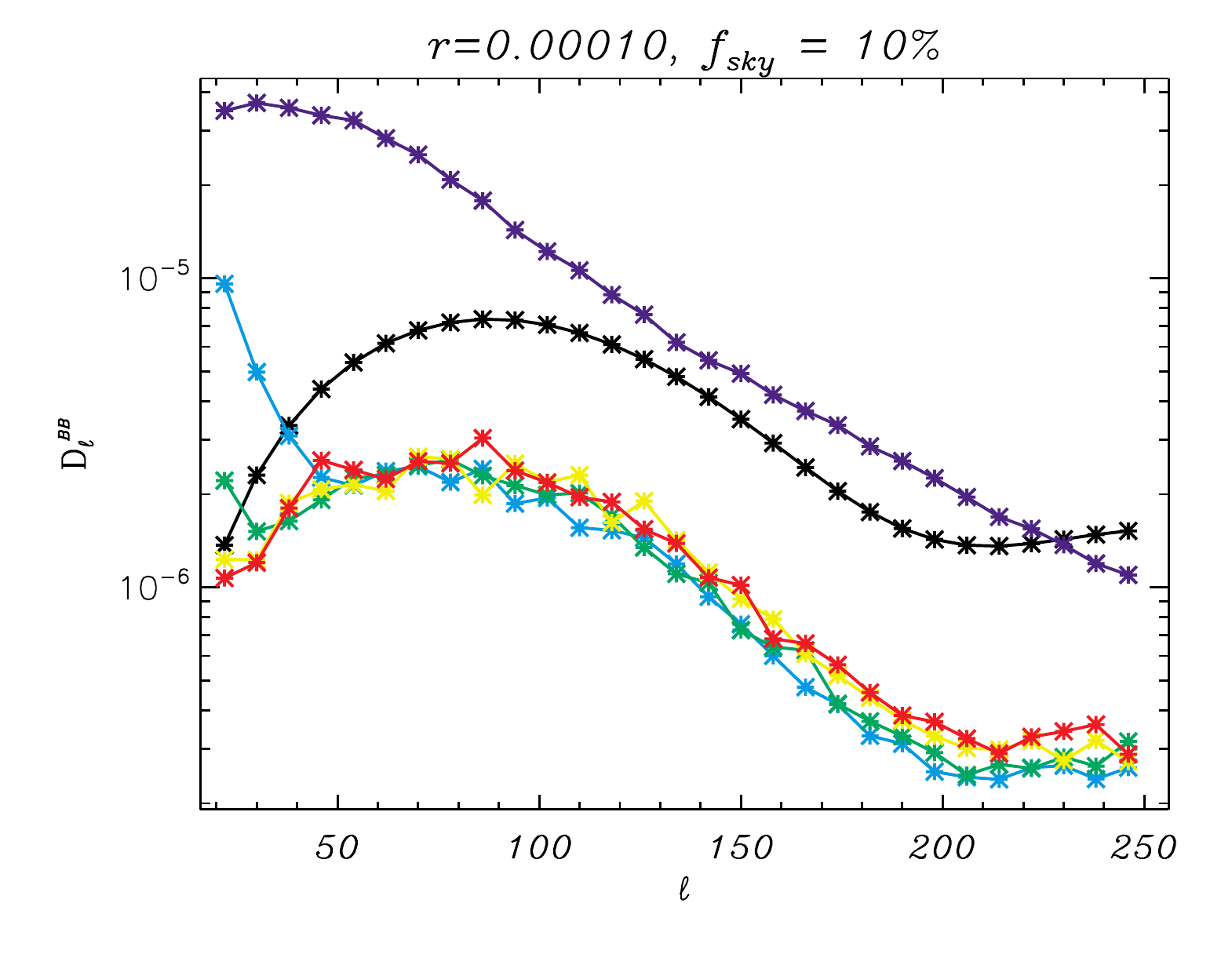}
  \includegraphics[width=0.32\textwidth]{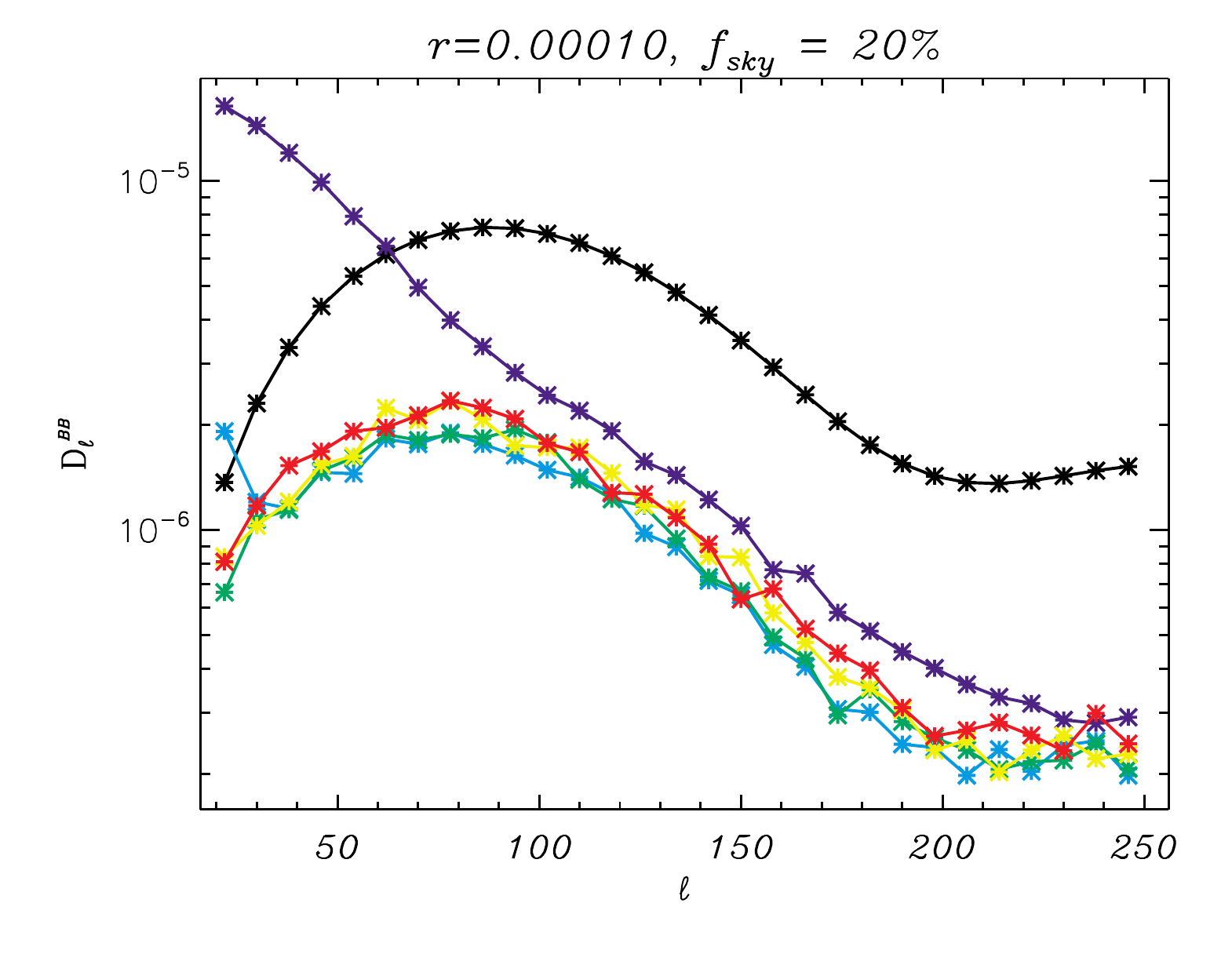}
  \caption{Similar to Figure~\ref{fig:test r5} but for $r=10^{-4}$.}
  \label{fig:test r4}
\end{figure}

\begin{figure}
  \centering
  \includegraphics[width=0.32\textwidth]{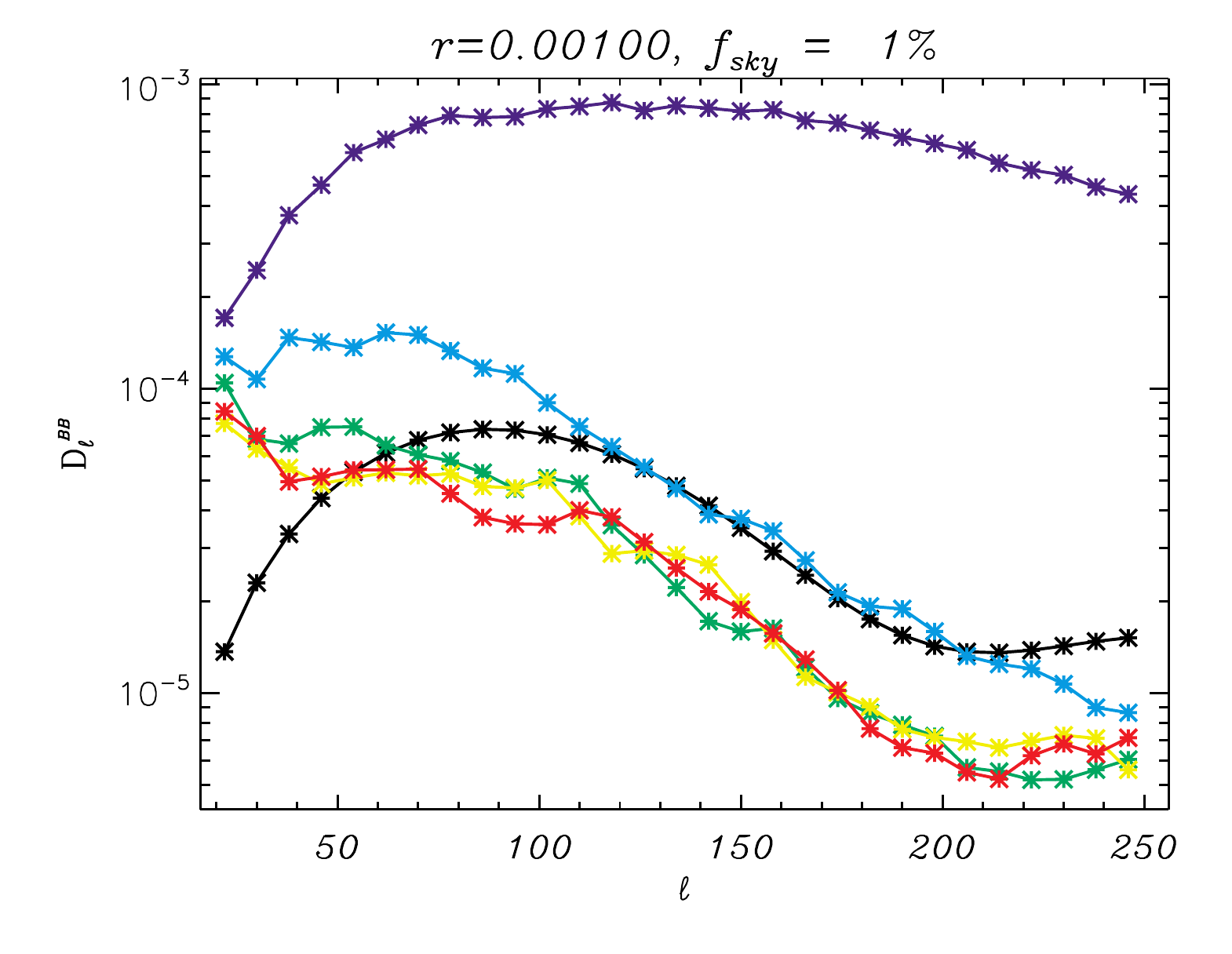}
  \includegraphics[width=0.32\textwidth]{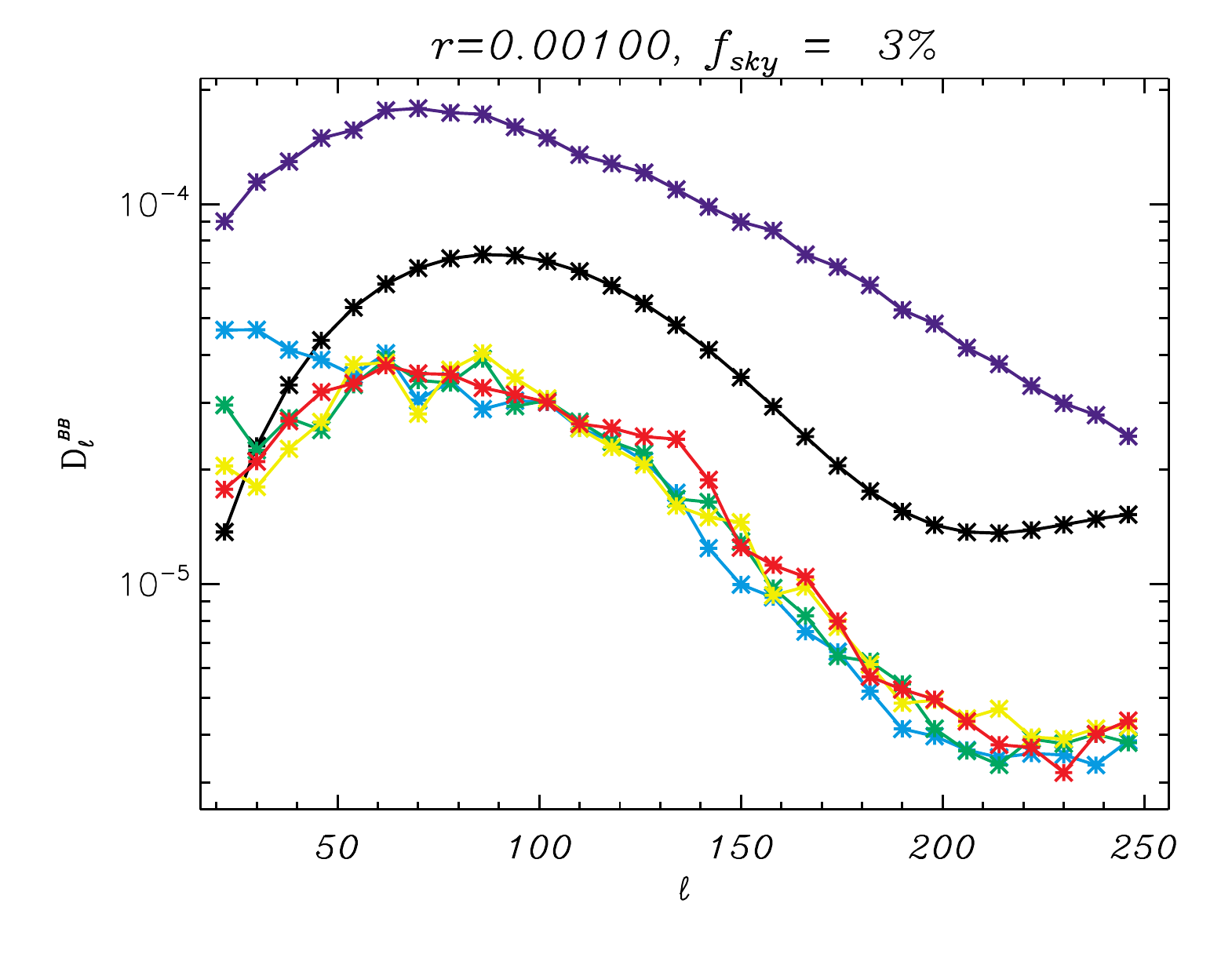}
  \includegraphics[width=0.32\textwidth]{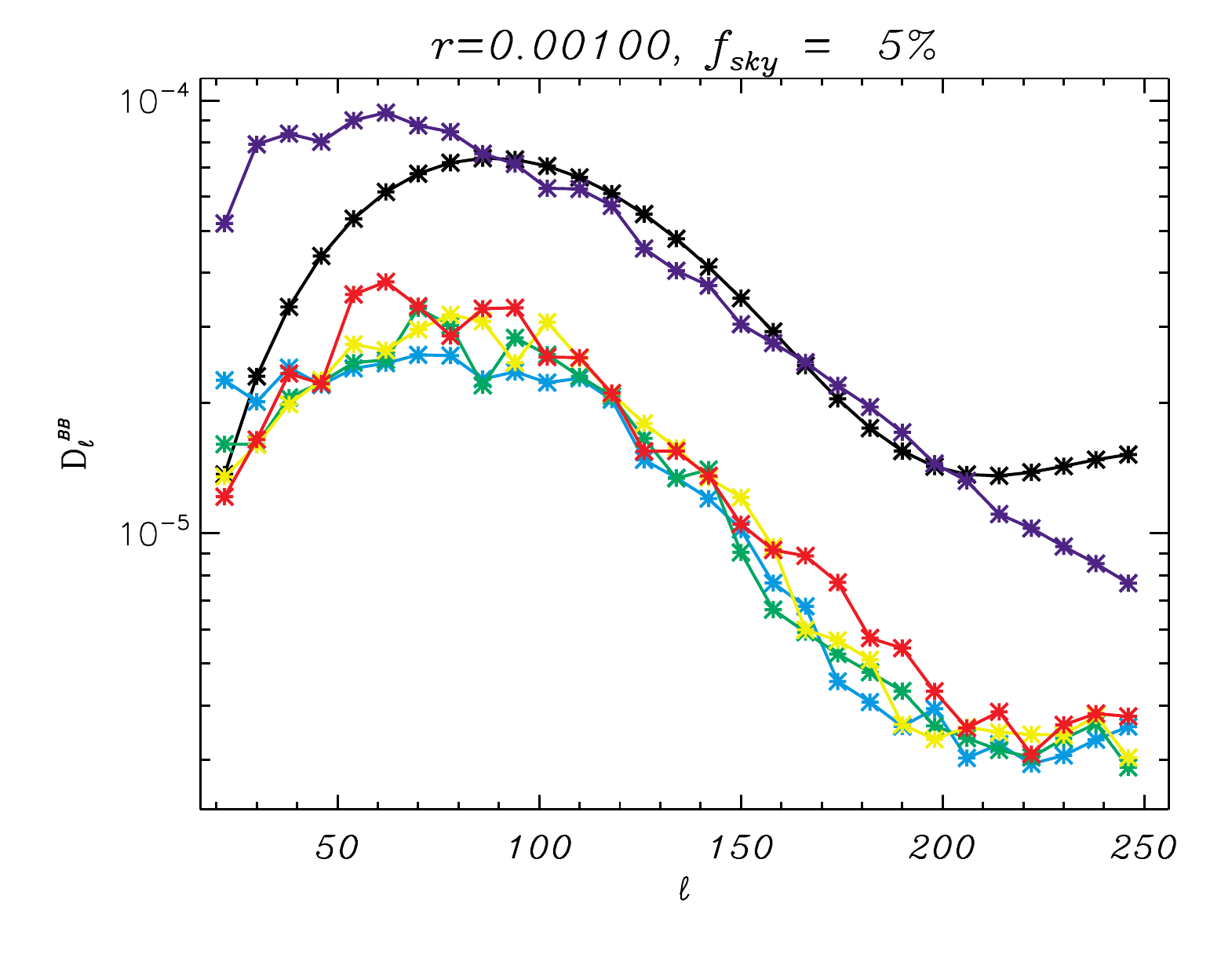}

  \includegraphics[width=0.32\textwidth]{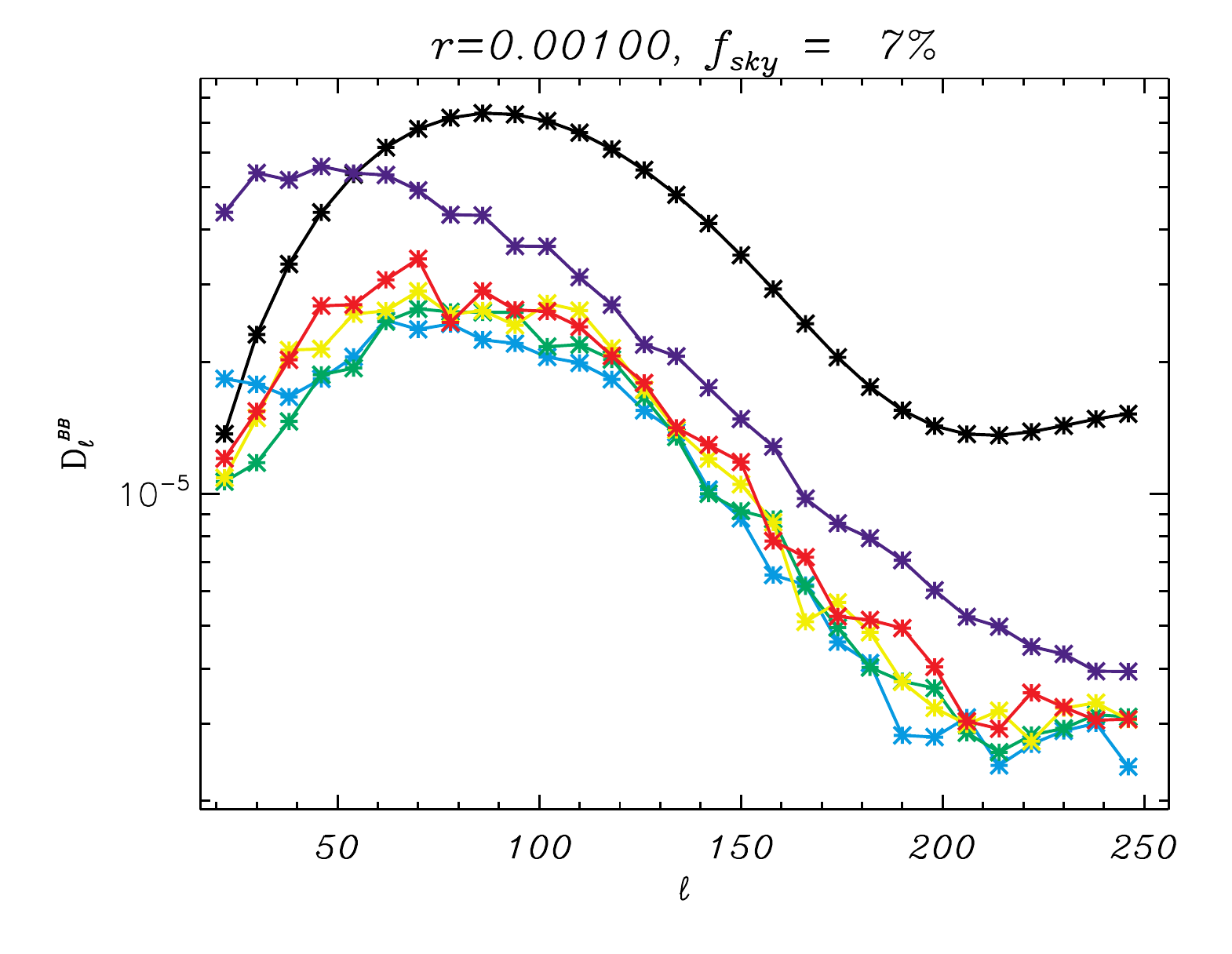}
  \includegraphics[width=0.32\textwidth]{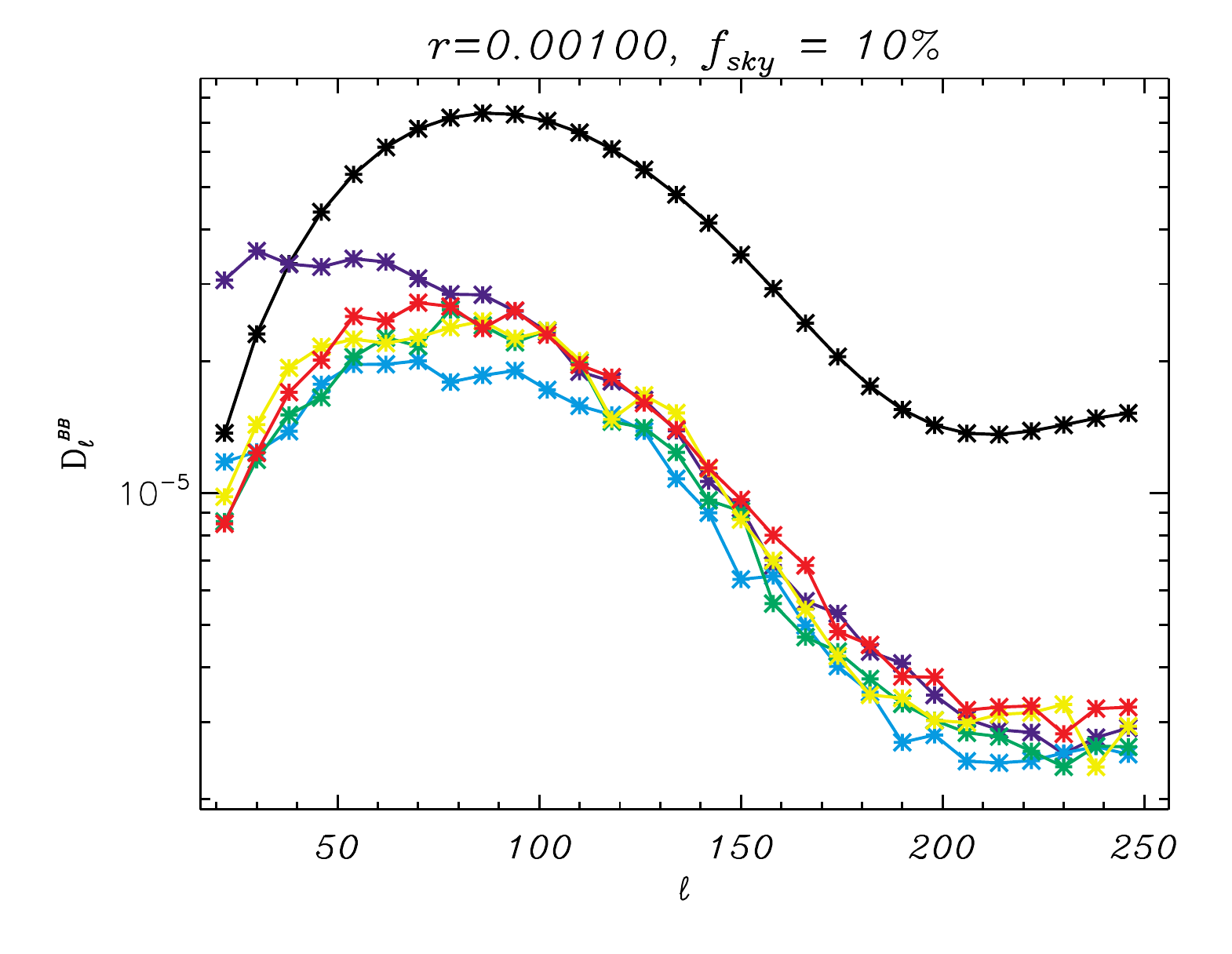}
  \includegraphics[width=0.32\textwidth]{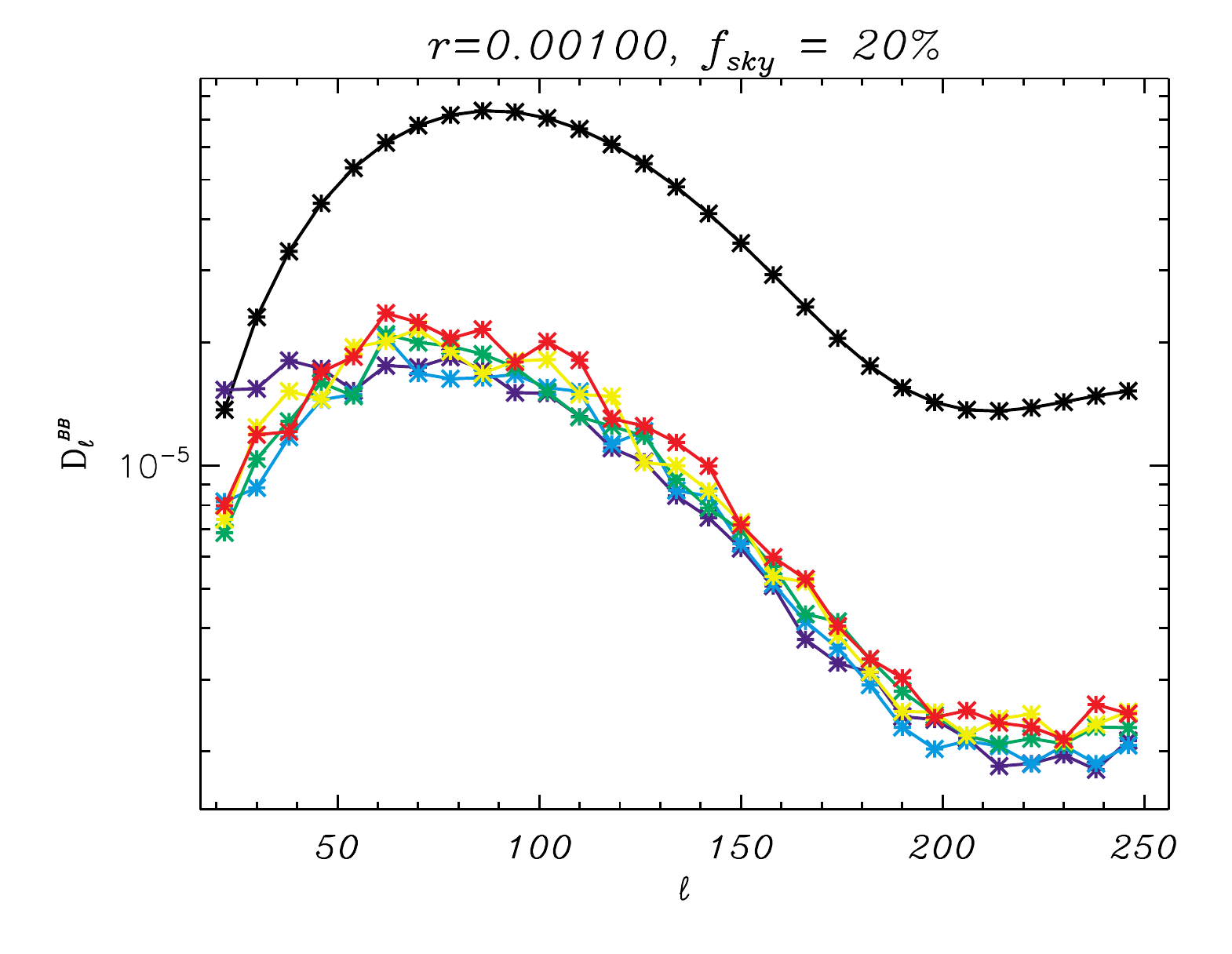}
  \caption{Similar to Figure~\ref{fig:test r5} but for $r=10^{-3}$.}
  \label{fig:test r3}
\end{figure}

\begin{figure}
  \centering
  \includegraphics[width=0.32\textwidth]{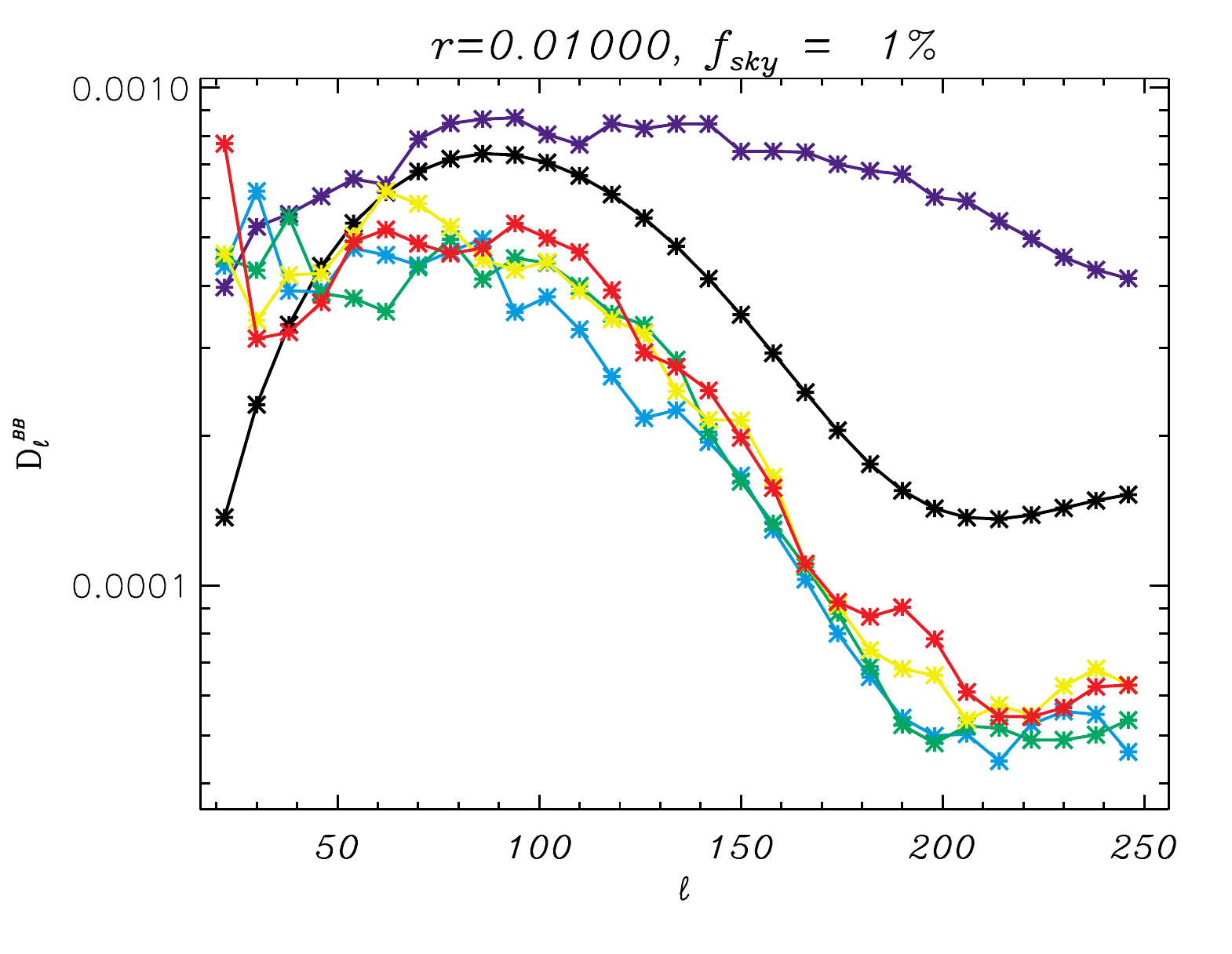}
  \includegraphics[width=0.32\textwidth]{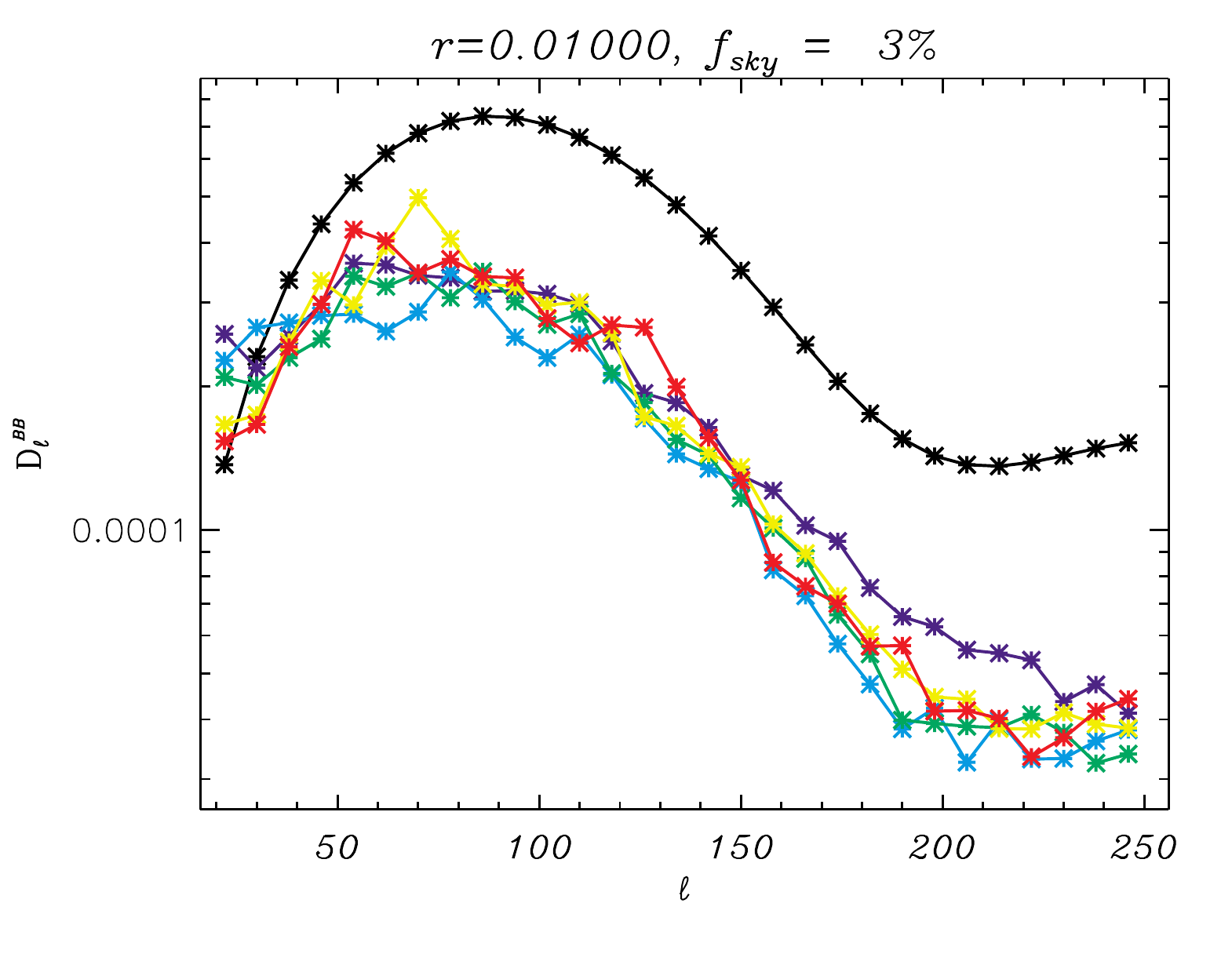}
  \includegraphics[width=0.32\textwidth]{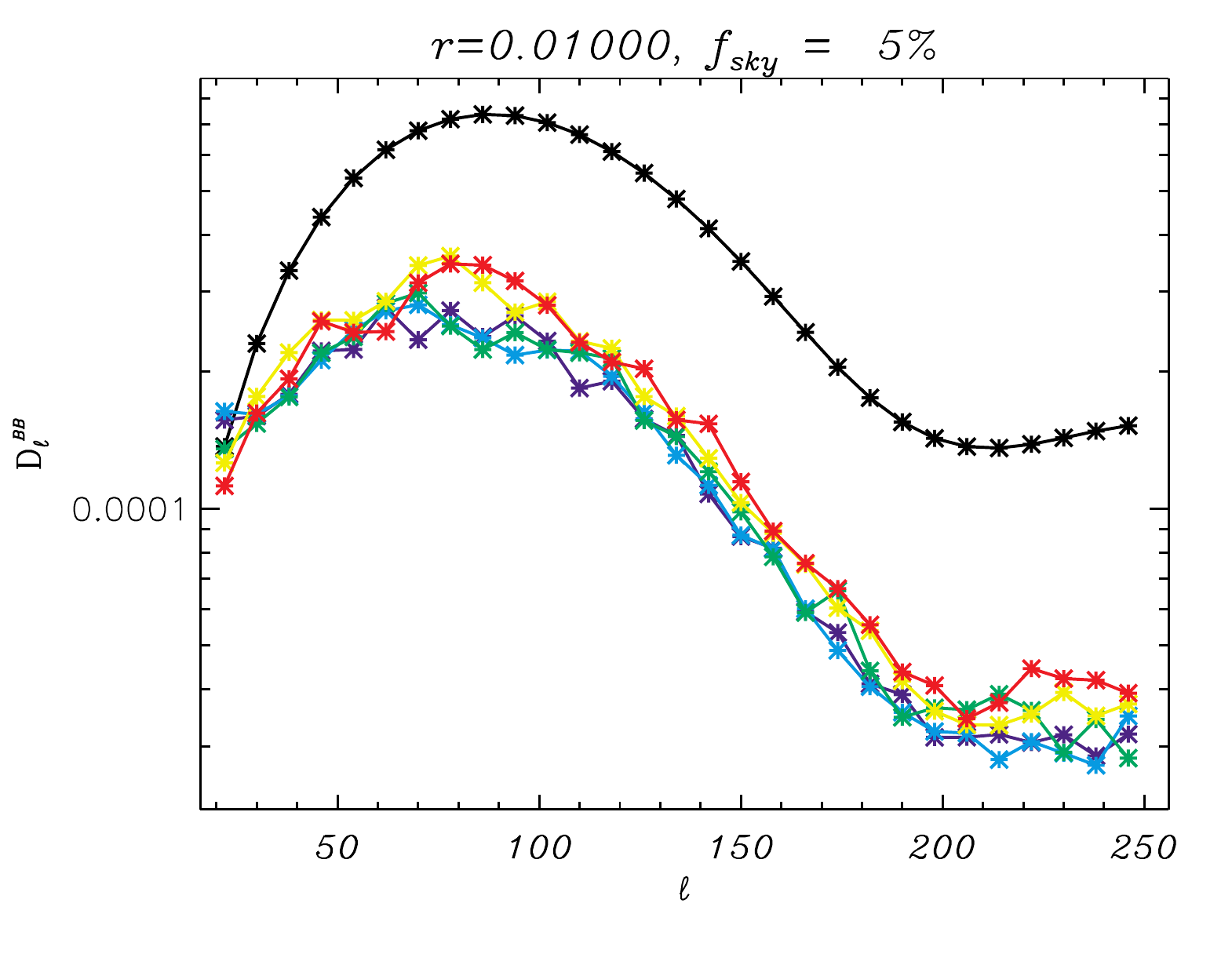}

  \includegraphics[width=0.32\textwidth]{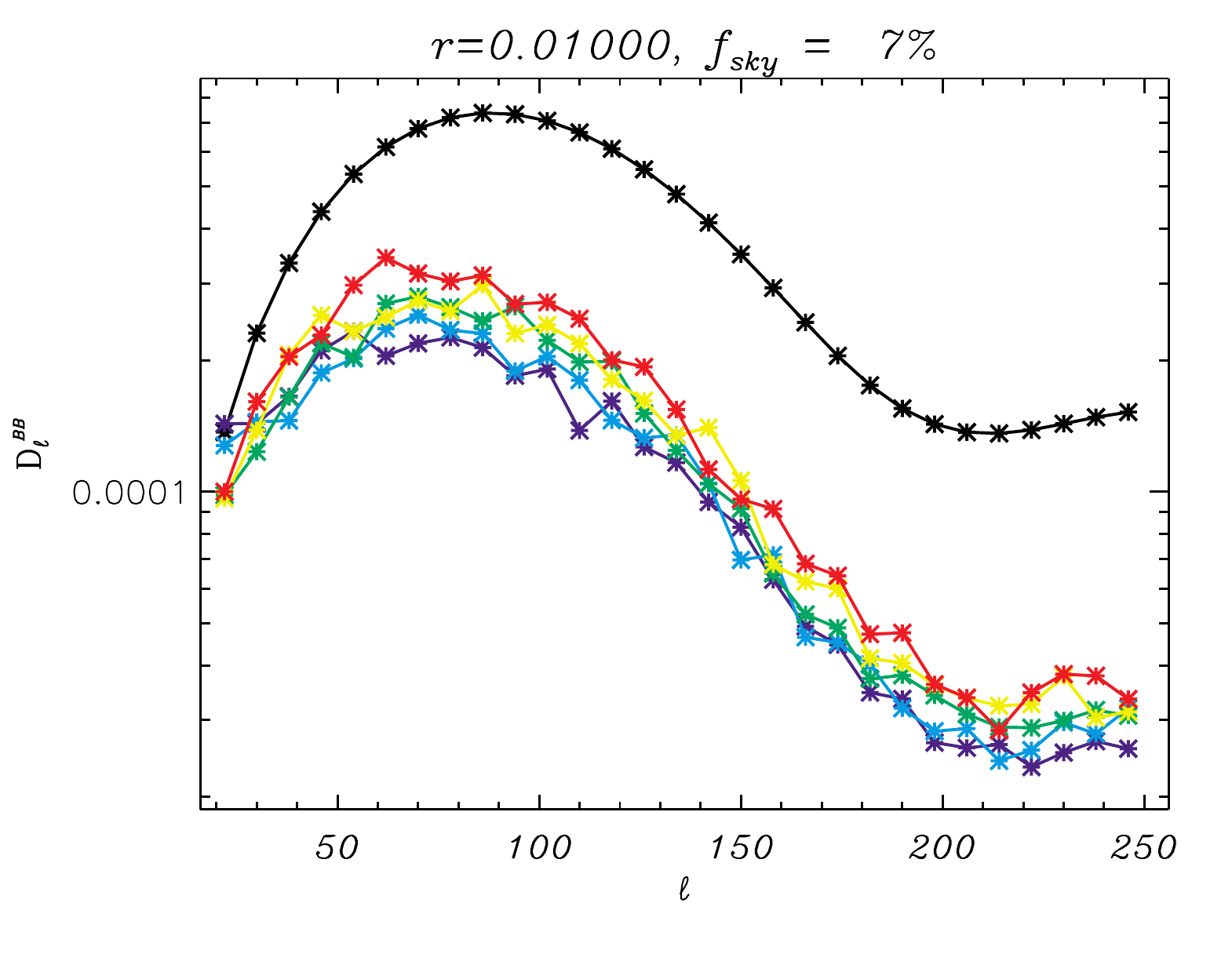}
  \includegraphics[width=0.32\textwidth]{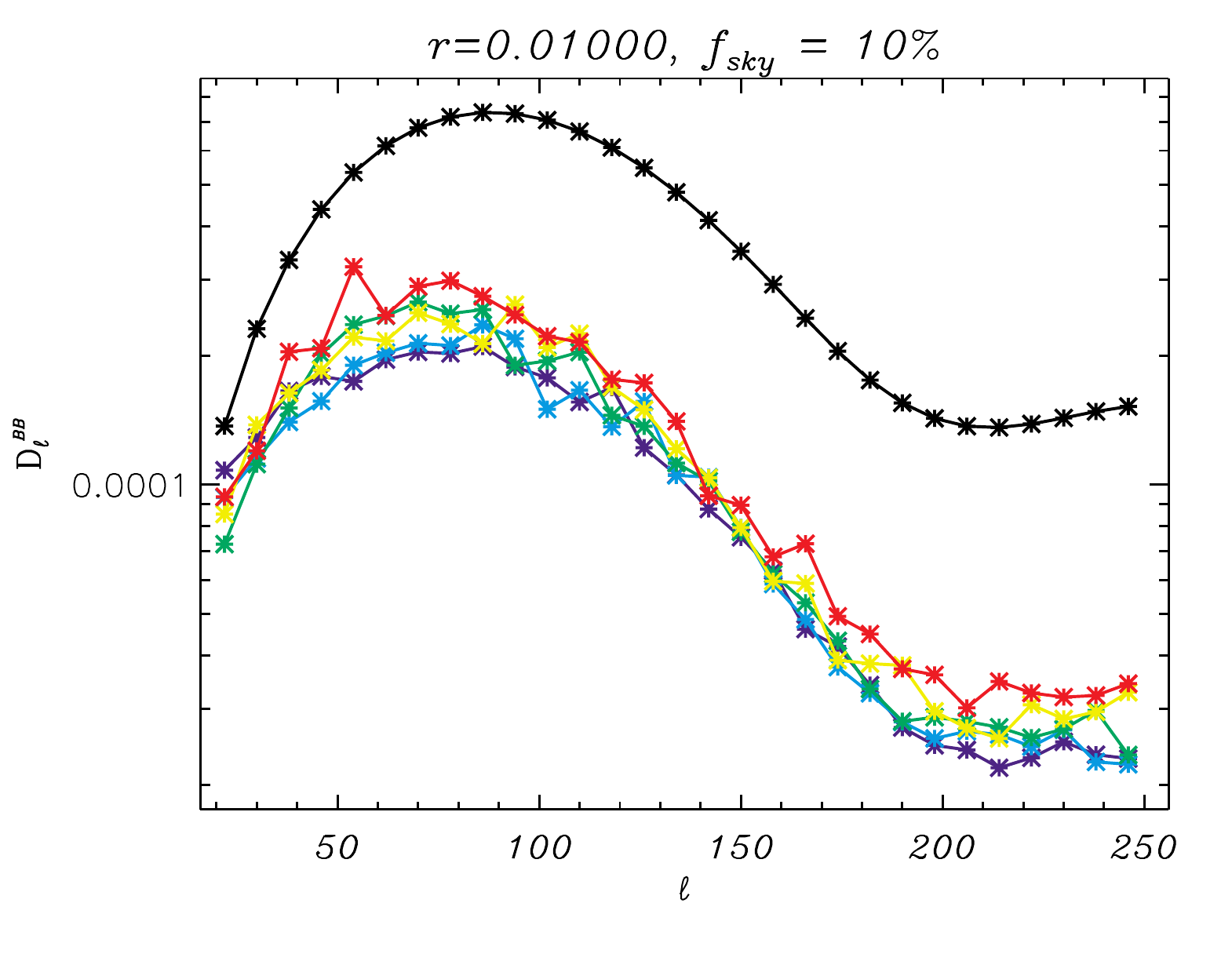}
  \includegraphics[width=0.32\textwidth]{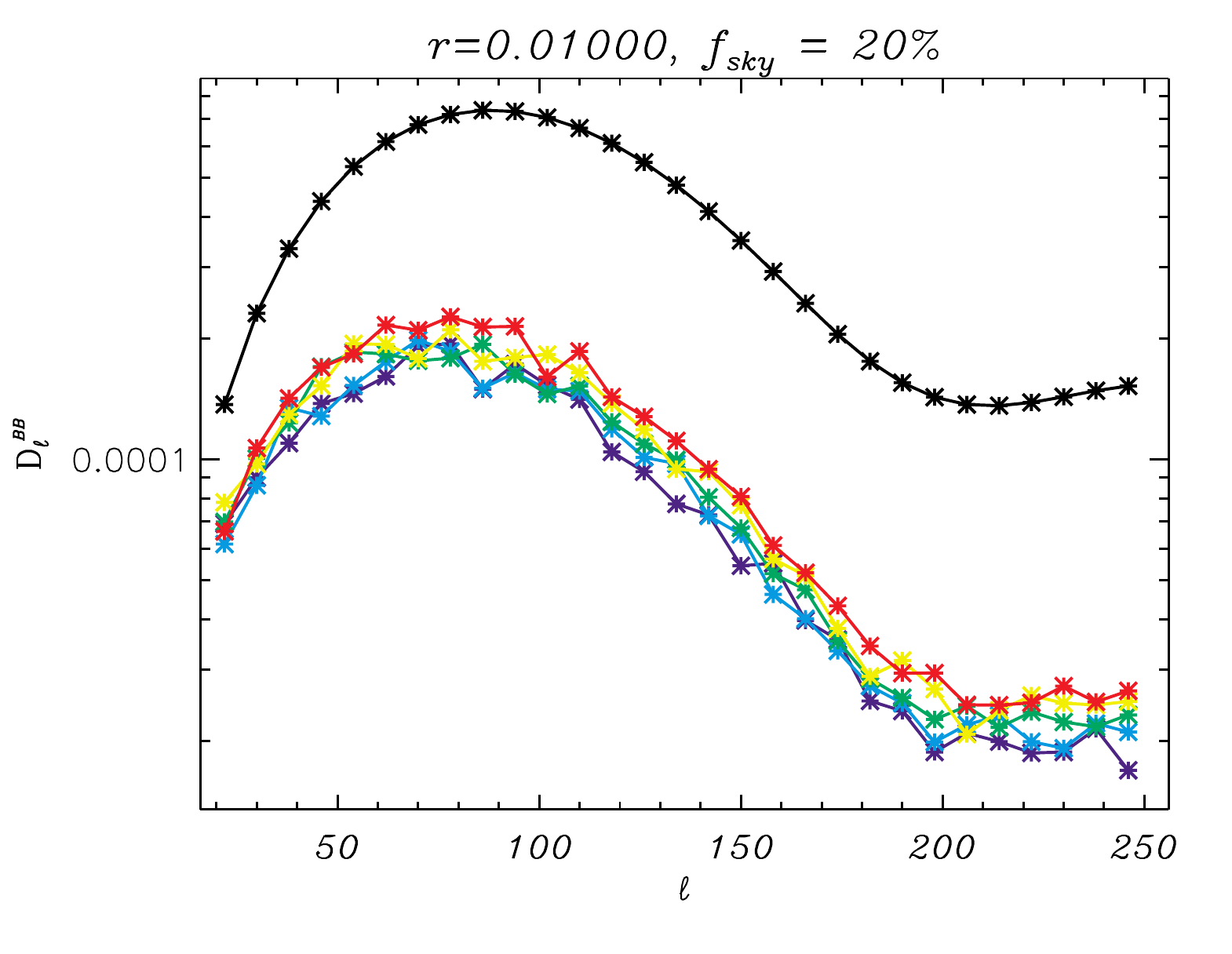}
  \caption{Similar to Figure~\ref{fig:test r5} but for $r=10^{-2}$.}
  \label{fig:test r2}
\end{figure}

\section{Summary and discussion}\label{sec:discuss}

In summary, the general solution of the leakage due to data missing is given
for integral transforms, and is used to estimate the maximum ability to detect
the CGWB through the CMB with incomplete sky coverage. The results are
presented in Figures~\ref{fig:test r5}--\ref{fig:test r2}, and for convenience
of understanding, a brief outlook based on these figures is summarized in
Table~\ref{tab:det}.

The price to use a prior estimator is to lose some generality. Theoretically
speaking, it is possible to avoid this price by a brute force or MCMC
exploration of all possible prior EE-spectra, and to find the one that gives
the smallest error bars. However, this means to pay another incredibly huge
price for the computational cost, which is not a good deal. In reality, since
an excellent EE-spectrum was already given by the Planck mission with full sky
surveys~\citep{2016A&A...594A...1P, 2018arXiv180706205P}, it is evidently a
good idea to use this EE-spectrum as the prior information in the problem of
the EB-leakage.

The BUE in this work has a great advantage that it can easily accommodate all
kinds of prior information/constraints, even if they are non-Gaussian or
non-analytic, e.g., realistic beam profile, systematics, lensing effects, etc.
The only requirement is that the corresponding effects can be simulated and
accommodated in $\{f_i(\bm{p})\}_{\mathcal{I}}$. An example of such
simulations is the set of Planck full focal plane (FFP)
simulations~\citep{2016A&A...594A..12P}. Meanwhile, any improvement of the
prior information will immediately help to improve the overall estimation.

Interestingly, as shown by Fig.~\ref{fig:cmp BUE}, for pure CMB signal, the
error bars of the recycling method~\citep{2018arXiv181104691L} are roughly
only $30\%$ bigger than the ideal error bars, thus the result of the recycling
method is an excellent approximation of the BUE, especially when computed at a
higher resolution. However, as mentioned above, this must be further tested if
non-Gaussian and non-analytic prior information is taken into account.

Finally, according to Table~\ref{tab:det}, to detect $r=10^{-3}\sim10^{-2}$,
it is recommended to use at least $f_{sky}=3\%$, and for
$r=10^{-5}\sim10^{-4}$, it is recommended to use $f_{sky}=10\%$ or more.

\Ack{

I sincerely thank Pavel Naselsky and James Creswell for valuable discussions,
and the anonymous referee for reading the manuscript carefully and giving
useful comments. This research has made use of the
\textsc{HEALPix}~\citep{2005ApJ...622..759G} package, and was partially funded
by the Danish National Research Foundation (DNRF) through establishment of the
Discovery Center and the Villum Fonden through the Deep Space project. Hao Liu
is also supported by the National Natural Science Foundation of China (Grants
No. 11653002, 11653003), the Strategic Priority Research Program of the CAS
(Grant No. XDB23020000) and the Youth Innovation Promotion Association, CAS.
\\

}

\appendix

\section{Least square fitting of multi-variants}\label{app:least square}

Given an ensemble of real data sets $\{H^i_\xi\}$ and an ensemble of
corresponding measurements $\{h^i_\xi\}$, where $i$ denotes the index of the
ensemble members, and $\xi$ denotes the index of data points within each
member. Assume an estimation of $H_\xi$ is given by
\begin{equation}\label{equ:basic assumption}
H_\xi\approx\widetilde{H}_\xi = \sum_{\xi'}\bm{M_{\xi\xi'}}h_{\xi'} ,
\end{equation}
then the error of estimation at each point is
\begin{align}\begin{split}\label{equ: error of est}
\delta_\xi^2 &= \sum_i{|H_\xi^i-\widetilde{H}^i_\xi|^2} 
= \sum_i{|H_\xi^i-\sum_{\xi'}\bm{M_{\xi\xi'}}h^i_{\xi'}|^2}.
\end{split}\end{align}

The minimal variance condition shows
\begin{align}\begin{split}\label{equ: min var condition}
\Delta=\sum_\xi \delta_\xi^2 = 
\sum_{i,\xi}{|H_\xi^i-\sum_{\xi'}\bm{M_{\xi\xi'}}h^i_{\xi'}|^2}
= \min,
\end{split}\end{align}
thus
\begin{align}\begin{split}\label{equ:patial deviation}
\frac{\partial \Delta}{\partial
\bm{M_{\xi\xi'}}} = 
&-2\sum_i{[H_\xi^i-\sum_{\xi''}{\bm{M}_{\xi\xi''}h_{\xi''}^i}]h^{i*} _ {\xi'}}=0 \\ 
= &\sum_i{H^i_\xi
h^{i*}_{\xi'}}-\sum_{\xi''}{\bm{M}_{\xi\xi''}\sum_i{h^i_{\xi''}h^{i*}_{\xi'}}},
\end{split}\end{align} 
where $*$ denotes the complex conjugate. We write
\begin{eqnarray} \label{equ:P-matrix}
\bm{P}_{\xi\xi'}=\sum_i{H^i_\xi h^{i*}_{\xi'}} \\ \nonumber
\bm{Q}_{\xi\xi'}=\sum_i{h^i_\xi h^{i*}_{\xi'}},
\end{eqnarray}
where $\bm{P}_{\xi\xi'}$ is the cross covariance matrix between $H_\xi$ and
$h_\xi$, and $\bm{Q}_{\xi\xi'}$ is the covariance matrix of $h_\xi$. Thus
equation~(\ref{equ:patial deviation}) becomes:
\begin{align}\begin{split}\label{equ:m-solution0}
\bm{P} = \bm{M}\bm{Q},
\end{split}\end{align}
and the coupling matrix that gives the BUE is
\begin{align}\begin{split}\label{equ:m-solution}
\bm{M} = \bm{P}\bm{Q^{-1}}.
\end{split}\end{align}
Note that this solution is also a general one: if $\bm{Q}$ is the covariance
matrix of the known variable, $\bm{P}$ is the cross covariance matrix between
the unknown and known variable, and the equation of estimation is linear, then
the BUE of the unknown variable is given by eq.~(\ref{equ:m-solution}).

\section{Equivalence between the Fisher estimator and the maximum likelihood
estimator}\label{app:QML and WMAP MCL}

Here we provide a step-by-step proof that, for a Gaussian isotropic signal like
the CMB, the Fisher estimator and the standard maximum likelihood estimator
(e.g.~\cite{2003ApJS..148..195V}) give identical results.

The likelihood of a multi-variant Gaussian field $\bm{X}=\{x_1,x_2,\cdots\}$
that can be described by a set of model parameters
$\bm{\Theta}=\{\theta_1,\theta_2,\cdots\}$ is:
\begin{equation}\label{equ:MCL}
L(\bm{X}|\bm{\Theta})\propto \frac{e^{-\frac{1}{2}\bm{X}^T C^{-1} \bm{X}}}{\sqrt{|C|}},
\end{equation}
where $|C|$ means to take the determinant, and $C_{ij}=\<x_ix_j\>$ is the
covariance matrix in the pixel domain, determined by $\bm{\Theta}$. The above
equation itself does not require isotropy, but by assuming isotropy,
$\bm{\Theta}$ can be simplified to $\bm{\Theta}\equiv C_\ell$, the angular
power spectrum, so the covariance matrix has the following form:
\begin{equation}\label{equ:C and legendre}
C_{ij} = \frac{1}{4\pi} \sum_\ell{(2l+1) W_\ell^2 C_\ell P_\ell[\cos{(\theta_{ij})}]},
\end{equation}
where $W_\ell$ is the beam profile, $P_\ell$ is the Legendre polynomial of
order $l$, and $\theta_{ij}$ is the angle between pixel $i$ and $j$.

With eq.~(\ref{equ:C and legendre}), the partial derivative of $C$ is:
\begin{equation}\label{equ:Pl}
\frac{\partial C}{\partial C_\ell} =\frac{1}{4\pi} (2\ell+1) 
W_\ell^2 P_\ell[\cos{(\theta_{ij})}] = \rm{const}
\end{equation}

The standard form of Fisher estimator for CMB is:
\begin{eqnarray}\label{equ:QML basic}
F_{\ell\ell'}&=&2\textbf{Tr}[C E^\ell C E^{\ell'}]\\ \nonumber
\widetilde{C_\ell} &=& F_{\ell\ell'}^{-1} y_{\ell'}
\end{eqnarray}
where $F_{\ell\ell'}$ is the Fisher matrix, and
\begin{equation}\label{equ:qml def}
E^\ell=\frac{1}{2} C^{-1} \frac{\partial C}{\partial C_\ell} C^{-1},
\,y_\ell=\bm{X}^TE^\ell\bm{X}
\end{equation}

Therefore, we should start from Equation~\ref{equ:MCL} and obtain
Equation~\ref{equ:QML basic}.

For Equation~\ref{equ:MCL} to get its maximum, we have
\begin{equation}\label{equ:mcl_partial}
\frac{ \partial{\log{L}}} {\partial{C_\ell}}=0,
\end{equation}
thus
\begin{equation}\label{equ:mcl_partial1}
-\bm{X}^T\frac{\partial C^-1}{\partial C_\ell}\bm{X}=
\frac{1}{|C|}\frac{\partial |C|}{\partial C_\ell}.
\end{equation}
Since $$\frac{ \partial} {\partial{C_\ell}}(C C^{-1})=0,$$ we have
\begin{equation}\label{equ:mcl_partial2}
\frac{ \partial{C}} {\partial{C_\ell}} C^{-1} + 
C\frac{ \partial{C^{-1}}} {\partial{C_\ell}}=0,
\end{equation}
which means
\begin{equation}\label{equ:mcl_partial3}
\frac{ \partial{C^{-1}}} {\partial{C_\ell}} = 
-C^{-1}\frac{ \partial{C}} {\partial{C_\ell}}C^{-1},
\end{equation}
so Equation~\ref{equ:mcl_partial1} becomes
\begin{equation}\label{equ:mcl_partial4}
\bm{X}^T C^{-1}\frac{ \partial{C}} {\partial{C_\ell}}C^{-1} \bm{X} = 
\frac{1}{|C|}\frac{\partial |C|}{\partial C_\ell}.
\end{equation}
With the definition of $y_\ell$, the above is shorted as
\begin{equation}\label{equ:yl}
2y_\ell= \frac{1}{|C|}\frac{\partial |C|}{\partial C_\ell}.
\end{equation}
For the right hand side, the variation of $C$ is
\begin{equation}\label{equ: dc}
C +\frac{ \partial{C}} {\partial{C_\ell}} dC_\ell = C k,
\end{equation}
so
\begin{equation}\label{equ:dc1}
k =I+C^{-1}\frac{ \partial{C}} {\partial{C_\ell}} dC_\ell,
\end{equation}
where $I$ is the unitary matrix. According to the definition of determinant,
\begin{equation}\label{equ:dc2}
d|C|=|C k|-|C|=|C|\,|k|-|C|=|C|\,\textbf{Tr}
\left(C^{-1}\frac{ \partial{C}} {\partial{C_\ell}}\right)dC_\ell.
\end{equation}
Thus
\begin{equation}\label{equ:dc3}
\frac{1}{|C|}\frac{ \partial{|C|}} {\partial{C_\ell}}=
\textbf{Tr}\left(C^{-1}\frac{ \partial{C}} {\partial{C_\ell}}\right)
\end{equation}
Substitute into Equation~\ref{equ:yl} gives
\begin{equation}\label{equ:yl_final}
y_\ell = \frac{1}{2}\textbf{Tr}\left(C^{-1}\frac{ \partial{C}} {\partial{C_\ell}}\right).
\end{equation}
For Gaussian field, $y_\ell$ is linear function of $C_\ell$, thus
\begin{align}\begin{split}\label{equ:get fll}
y_\ell 
&= \sum_{\ell'} \frac{ \partial y_\ell } {\partial{C_{\ell'}}} C_{\ell'} \\ 
\frac{ \partial y_\ell}{\partial C_{\ell'}} 
&= F_{\ell\ell'} = \frac{1}{2}\frac{\partial\textbf{Tr}
(C^{-1}\frac{ \partial{C}} {\partial{C_\ell}} )}{\partial C_{\ell'}}.
\end{split}\end{align}
Using Equation~\ref{equ:mcl_partial2}, and considering that $
\partial{C}/\partial{C_\ell}$ is constant and $C$ is symmetric, we get
\begin{align}\begin{split}\label{equ:dyl1}
F_{\ell\ell'} &=\frac{ \partial y_\ell}{\partial C_{\ell'}} \\\nonumber
&= \frac{1}{2}\textbf{Tr}\left[\frac{\partial C^{-1}}{\partial C_{\ell'}}
\frac{ \partial{C}} {\partial{C_\ell}} \right]\\\nonumber
&=\frac{1}{2}\textbf{Tr}\left[C^{-1}\frac{\partial C }{\partial C_{\ell'}}C^{-1}
\frac{\partial C }{\partial C_{\ell}}\right]\\\nonumber
&=2\textbf{Tr}[C E^\ell C E^{\ell'}],
\end{split}\end{align}
which returns exactly to Equation~\ref{equ:QML basic}. Therefore, for a Gaussian
isotropic signal like the CMB, the Fisher estimator and maximum likelihood
estimator are identical.

\providecommand{\href}[2]{#2}\begingroup\raggedright\endgroup

\end{document}